\documentclass[10pt,preprint,showpacs,preprintnumbers]{revtex4}
\usepackage{amssymb,amsthm,epsfig,graphics,graphicx,subfigure}

\renewcommand{\d}{\mathrm{d}}
\renewcommand{\t}{\mathrm{Ta}}
\renewcommand{\c}{\mathrm{corr}}

\begin{document}

\title{Lifetime of micrometer-sized drops of oil pressed by buoyancy against a planar interface}

\author{Clara Rojas}
\email{clararoj@gmail.com}
\author{Germ\'an Urbina-Villalba}
\author{M\'aximo Garc\'ia-Sucre}

\affiliation{Centro de Estudios Interdisciplinarios de la F\'isica IVIC Apdo 20632, Caracas 1020A, Venezuela}

\date{\today}

\begin{abstract}
Emulsion Stability Simulations (ESS) are used to estimate the coalescence time of one drop of hexadecane pressed by buoyancy against a planar water/hexadecane interface. In the present simulations the homophase is represented by a big drop of oil at least $500$ times larger than the approaching drop ($1\,\mu$m to $10\,\mu$m). Both deformable and non-deformable drops are considered along with six different diffusion tensors. In each case van der Waals, electrostatic, steric and buoyancy forces are taken into account. The coalescence times are estimated as the average of $1000$ random walks. It is found that the repulsive potential barrier has a significant influence in the results. The experimental data can only be reproduced assuming negligible repulsive barriers, as well as non-deformable drops that move with a combination of Stokes and Taylor tensors as they approach the interface. 
\end{abstract}


\maketitle

\newpage
\section{Introduction}

Nowadays it is not yet possible to measure the coalescence time between two small drops of oil suspended in quiescent water. In an approximation to the problem four different experimental techniques are often used to determine: (a) the lifetime of thin films of macroscopic radii \cite{velikov:1997}, (b) the nature of the forces between two fixed drops (Atomic Force Microscopy) \cite{dagastine:2006,webber:2008}, (c) the dynamic interaction forces between two particles during flow \cite{casanova:2000}, and (d) the coalescence time of oil droplets pressed against a planar oil-water interface \cite{deshiikan:1995,dickinson:1988,basheva:1999}. The first technique allowed identifying the stages of the coalescence process between deformable interfaces. The second showed significant differences between the behavior of films and droplets during coalescence. The third method outlined the importance of the Brownian movement in the scattering of particles of micron size covered by $\beta$-casein. The fourth technique evidenced an interesting dependence of the coalescence time with the radius of the approaching drop. 

Using optical microscopy Dickinson {\it et al.} \cite{dickinson:1988} reported the coalescence time of micron-size ($1$ to $5\,\mu$m) droplets of hexadecane pressed by buoyancy against a planar interface. For this purpose, a small drop of oil was released in an aqueous protein solution, just below a planar water/hexadecane interface. The coalescence time between the droplet and the interface was determined by direct observation using optical microscopy. Proteins ($\beta$-casein, $\kappa$-casein and lysozyme) were used in order to guarantee the immobility of the surfactant layer at the oil/water interface. However, the surfactant concentration was kept low enough in order to guarantee the coalescence of the drops. It was found that independently of the protein used, the coalescence time decreased with the increase of the radius of the droplets. 

Ivanov and Kralchevsky \cite{ivanov:1997} studied the hydrodynamic flow between a sphere and a planar interface, emphasizing on the effect of the drop deformability on the coalescence time. They predicted a minimum of the coalescence time as a function of the drop radius for the case of tangentially immobile interfaces and a fixed driving force. Basheva {\it et al.} \cite{basheva:1999} studied the asymptotic limits of this size dependence both theoretically and experimentally. Using an experimental set up similar to the one of Dickinson {\it et al.}, they reported a decrease of the coalescence time of soybean oil drops ($2\,\mu\textnormal{m} \leq r_i \leq 100\,\mu$m) pressed against a planar water/oil interface. For these experiments, Bovine Serum Albumin (BSA) was used as a surfactant. 

In order to explain their experimental results, Basheva {\it et al.} \cite{basheva:1999} argued that when a drop is released in a bulk liquid, it moves according to Stokes law until it approaches the interface to a sufficiently small distance. Then it markedly decelerates due to the increase of the viscous friction in the remaining gap. Hence, the authors defined the lifetime of the drop at the interface, $\tau$, as the time elapsed between the moment in which the drop start to move slowly until it disappears by coalescing with the large homophase:  

\begin{equation}
 \tau=\int^{h_{\mathrm{ini}}}_{h_\mathrm{crit}} \frac{\d h}{v(h)},
\label{lifetime}
\end{equation}
where $v(h)$ is the velocity of thinning of the liquid film between the drop and the interface, $h$ is the closest distance of approach between their surfaces, $h_\mathrm{ini}$ is the gap width at which the thinning begins, and $h_\mathrm{crit}$, the minimum distance that can be attained (critical thickness) before the film breaks and coalescence occurs.  
If the interfaces of the drop and the planar boundary are tangentially immobile, and if the drop keeps its spherical shape until contact, then the expression of Taylor for the drainage of the intervening film between hard spheres can be used  \cite{taylor:1923} 

\begin{equation}
 v(h)=v_{\t} = \frac{2hF}{3\pi\eta r_*^2}, \quad r_*=\frac{2r_i r_j}{r_i+r_j}.
\label{two-spheres}
\end{equation}
Here $F$ is the external driving force, $\eta$ is the dynamic viscosity of the external liquid, $r_i$ is the radius of the small droplet and $r_j$ is the radios of the large drop.
Making $r_j\rightarrow \infty$:

\begin{equation}
 v_{\t} = \frac{hF}{6\pi\eta r_i^2}.
\label{sphere-flat}
\end{equation}

Basheva {\it et al.} assumed that the velocity of the approaching drop could be expressed as a combination of the velocities predicted by the tensors of Taylor and Stokes:

\begin{equation}
 \frac{1}{v}=\frac{1}{v_\mathrm{St}}+\frac{1}{v_\t},
\label{st+ta}
\end{equation}
where $v_\t$ is given by Eq. (\ref{sphere-flat}) and $v_\mathrm{St}=F/6\pi\eta r_i$ refers to the Stokes law for motion of a sphere in an unbounded liquid. 

Substituting Eq. (\ref{st+ta}) in Eq. (\ref{lifetime}), and using the expression for the buoyancy force:

\begin{equation}
F =\frac{4}{3}\pi r_i^3 \Delta\rho g,
\label{bouyancy}
\end{equation}
where $\Delta\rho$ is the density difference and $g$ is the gravity,
an inverse dependence between the lifetime of the drop at the planar interface, and its radius is obtained:

\begin{equation}
 \tau=\frac{9\eta}{2\Delta\rho g}\frac{1}{r_i}\left[\log\left(\frac{h_\mathrm{ini}}{h_\mathrm{crit}}\right)+\frac{h_\mathrm{ini}-h_\mathrm{crit}}{r_i}\right].
\label{Stokes-Taylor}
\end{equation}

According to Eq. (\ref{Stokes-Taylor}), the lifetime diminishes with the increase of the droplet radius $r_i$. This trend is just the opposite to the one found for drops of hundreds of microns (millimeter size). In the latter case the drops deform close to the interface, producing a circular film. This film drains according to Reynolds law, producing an increase in the lifetime of the drops as a function of their radii \cite{basheva:1999}.

It is remarkable that the experimental data on the coalescence of a drop with a planar interface follows Taylor law, since this expression was initially deduced for the case of two colliding spheres. However, it is well known that the radius of curvature of a drop increases with its size, approaching asymptotically the radius of curvature of a planar interface. Conversely, the behavior of the drop/interface system could be simulated using two drops of very distinct sizes. One big drop fixed in space and sufficiently large to be regarded as a planar interface by the approaching droplet, and one moving drop of micrometer size rising as a consequence of the buoyancy force. 

In this work the algorithm of Emulsion Stability Simulations (ESS) is used to reproduce the data of Dickinson {\it et al.} regarding drops between $1$ and $10\,\mu$m.  It will be shown that the simulations fit the experimental data if the approaching drop is assumed to be a non-deformable sphere, moving with the diffusion tensors of Stokes or Taylor depending on the distance of approach. It is also demonstrated that the use of a truncated spheres as a model of deformable drops does not fit the experimental data for these range of sizes, independently of the tensor employed. 

The article is structured as follows: In Sec. \ref{emulsion} an overview of Emulsion Stability Simulations is presented, in Sec. \ref{computational} a description of the calculations is given along with some technical details, in Sec. \ref{parametrization} the parametrization of the potentials is explained, and some preliminary results concerning the form of the potentials are presented. In Sec. \ref{results} we show the results of the calculations and Sec. \ref{conclusions} presents the conclusions.

\section{Emulsion Stability Simulations}
\label{emulsion}

Emulsion Stability Simulations are based on the algorithm of Brownian Dynamics published by Ermak and McCammon \cite{ermak:1978,urbina:2009a,toro:2009}. 
If the divergence of the diffusion tensor is assumed to be negligible, the displacement of particle $i$, $\mathbf{r}_i(t+\Delta t)-\mathbf{r}_i(t)$ is equal to:

\begin{equation}
\mathbf{r}_i(t+\Delta t)=\mathbf{r}_i(t)+\frac{D_{i}\mathbf{F}_i}{k T}\Delta t+\mathbf{R},
\label{brownian}
\end{equation}
where the second term on the right hand side accounts for the effect of conservative forces on the particle movement, $D_{i}$ is an effective diffusion constant (tensor) of particle $i$, $\mathbf{F}_i$ is the total force acting on $i$, $k$ is the Boltzmann constant, $T$ the temperature, $\Delta t$ the time step, and $\mathbf{R}$ is a random term representing the Brownian motion of the particle.

The diffusion constant is equal to  $D_{i}=D_0 f_{\c}^{(1)}f_{\c}^{(2)}$ where $D_0=kT/6\pi\eta r_i$ is the diffusion constant of Stokes. The first correction term $f_{\c}^{(1)}$, takes into account those factors that change the expression of the diffusion constant at infinite dilution \cite{padstow:1989}. The second correction term $f_{\c}^{(2)}$ takes into account the hydrodynamic interactions between the particles caused by the movement of the surrounding liquid as the particles advance \cite{urbina:2009a}.

In the most common case, a calculation begins distributing a set of oil drops in a cubic box of side length $L$. It is assumed that the molecules of oil mainly determine the van der Waals interaction between the particles. Instead, the repulsive interactions depend on the amount and chemical nature of the surfactant molecules adsorbed to the interface of the drops. The program has several routines for apportioning surfactant molecules amongst the drops. Once the surfactant has been distributed, the surface properties of the drops (like charge, interfacial tension, etc.) can be calculated. Then, the diffusion constant and interaction forces can be computed and the drops moved using Eq. (\ref{brownian}). At every time step, the program checks for the coalescence of drops. In the case of non-deformable drops, coalescence occurs whenever the distance of separation between the centers of mass, $r_{ij}$, is smaller than the sum of the radii of the drops. When this happens, a new drop is created at the center of mass of the coalescing particles. 

The present version of the code can either simulate the behavior of non-deformable or deformable drops. In both cases the particles follow the same equation of motion, Eq. (\ref{brownian}), but the analytical form of the diffusion tensor and the interaction forces change. 

If the mode of deformable droplets is selected, it is assumed that the deformation of the drops occurs independently of the energy required for this process. Due to its simplicity, the model of truncated spheres is used \cite{ivanov:1999,danov:1993b,danov:1993a}. According to this model, three regions of approach between two particles are defined:

\begin{description}
\item{Region I:} If the distance of separation between the centers of mass of the drops, $r_{ij}$, is larger than $r_i+r_j+h_{\mathrm{ini}}$, where $h_{\mathrm{ini}}$ stands for the initial distance of deformation, the drops maintain their spherical shape.  Consequently, the radius of the liquid film between the flocculating drops is zero, $r_\mathrm{f}=0$.

\item{Region II:} The drops change their shape from spheres to truncated spheroids. This region covers the range of distances between the beginning of the deformation $r_\mathrm{f}\neq 0$, and the attainment of the maximum film radius: $r_\mathrm{f}=r_{\mathrm{fmax}}=\sqrt{r_ih_{\mathrm{ini}}}$. In this zone, the close distance of separation between the surfaces of the drops is assumed to be constant $h=h_{\mathrm{ini}}$ \cite{ivanov:1999}, and:

\vspace{-0.6cm}
\begin{eqnarray}
\label{def1}
\nonumber
 h_{\mathrm{ini}}&+&\left(\sqrt{r_i^2-r_ih_{\mathrm{ini}}}+\sqrt{r_j^2-r_ih_{\mathrm{ini}}}\right)\\
&<& r_{ij}< r_i+r_j+h_{\mathrm{ini}},\\
\label{def2}
 r_\mathrm{f}&=&\sqrt{r_i^2-\left[\frac{r_i(r_{ij}-h_{\mathrm{ini}})}{r_i+r_j}\right]^2}.
\end{eqnarray}

\item{Region III:} The film radius already attained its maximum value, $r_\mathrm{f}=r_{\mathrm{fmax}}=\sqrt{r_ih_{\mathrm{ini}}}$, and the intervening liquid drains until it takes a critical distance of approach:

\vspace{-0.6cm}
\begin{eqnarray}
\label{def3}
h_\mathrm{crit}&+&\left(\sqrt{r_i^2-r_ih_{\mathrm{ini}}}+\sqrt{r_j^2-r_ih_{\mathrm{ini}}}\right)\\
\nonumber
&<& r_{ij}\\
\nonumber
&<& h_{\mathrm{ini}}+\left(\sqrt{r_i^2-r_ih_{\mathrm{ini}}}+\sqrt{r_j^2-r_ih_{\mathrm{ini}}}\right),\\
\label{def4}
 h&=&r_{ij}-\left(\sqrt{r_i^2-r_\mathrm{f}^2}+\sqrt{r_j^2-r_\mathrm{f}^2}\right).
\end{eqnarray}

\end{description}

Accurate estimation of the initial distance of deformation $h_\mathrm{ini}$, is very difficult since it results from a balance between hydrodynamic and interaction forces. The combination of the movement of the particles with the numerical solution of the exact formula of $h_\mathrm{ini}$ \cite{danov:1993a} is too demanding in terms of computational resources. Hence, we fitted the curves obtained in Ref. \cite{danov:1993a} with a polynomial expression for the approximate estimation of $h_\mathrm{ini}(r_{i},\gamma)$: 

\begin{eqnarray}
\nonumber
 h_\mathrm{ini}&=&\left[1.2932\times 10^8-8.6475\times 10^{-9}\right.\\
\nonumber
 &\times&\left.\exp(-r_i/1.8222\times 10^{-6}\right]\\
 &\times&\frac{3.3253+5.9804\exp(-\gamma/0.00402)}{3.3253+5.9804\exp(-10^{-3}/0.00402)},
\label{hini}
\end{eqnarray}
where $\gamma$ is the interfacial tension.
For the value of $h_\mathrm{crit}$, the expression published by Scheludko and others is used \cite{scheludko:1967,ivanov:1970,manev:2005a,manev:2005b}

\begin{equation}
h_\mathrm{crit}=\left(\frac{A_H A_\mathrm{crit}}{128\gamma}\right)^{1/4}, 
\end{equation}
where $A_\mathrm{crit}= r_\mathrm{f}/10$ and $A_H$ is the Hamaker constant.

Even in the mode of deformable drops, the particles behave as spheres if $r_{ij}>r_i+r_j+h_\mathrm{ini}$. This means that the potential of interaction and diffusion constant correspond to the ones of spherical particles within Region I. At $h=h_\mathrm{ini}$, the code calculates the dimensions of truncated spheres which are compatible with the actual distance of separation between the centers of mass of the spherical drops ($r_{ij}<r_i+r_j+h_\mathrm{ini}$). In this case, the expressions of the potentials corresponding to two truncated spheres are employed. Different expressions for the interaction potentials of truncated spheres are available from the literature \cite{danov:1993a}. They include, van der Waals, electrostatic, etc, see Table \ref{potentials}. Those potentials are expressed in terms of the width of the film and the particle radius. Use of Eqs. (\ref{def1})-(\ref{def4}), allows algebraic differentiation of the potentials in terms of $r_{ij}$. As a result it is possible to obtain analytical expressions for the force.  Those expressions can be evaluated using the value of the film width and film radius corresponding to each region of approach. 

It is important to remark that a pair of drops does not necessarily move sequentially between regions I, II and III. The movement of the particles described by Eq. (\ref{brownian}) may lead a couple of particles to go into the deformation region and back as a consequence of the random movement of the particles and/or their interaction forces. The program uses Eq. (\ref{brownian}) to move each particle separately, although the analytical form of the forces and diffusion constants depend on the relative distance between the particles.

Table \ref{potentials} shows the analytical form of the potentials employed in the present calculations. Notice that the geometrical part of the van der Waals and electrostatic potential changes as a function of deformation. Additionally, two new potentials appear during the evolution of the film (Region II). They take into account: (a) the surface deformation energy coming from the increase of interfacial area as the spherical drops turn into truncated spheres, and (b) the bending elasticity potential related to the curvature of the interface \cite{ivanov:1999}. These two potentials change with the interparticle distance during the formation of the film (Region II), but reaches a constant value once a maximum film radius has been attained. Hence, they do not contribute to the value of the force in Region III where Eqs. (\ref{def3}-\ref{def4}) hold.

Notice that in the present calculations the same form of the steric potential for spherical and deformable droplets is used (Table \ref{potentials}). This was done on purpose for several reasons. First, it avoids some anomalies in the behavior of the potential that might occur when the volume of overlap changes as a consequence of the sudden transition between spherical particles and truncated spheroids (Regions I and II). Second, it simplifies the parametrization of the steric potential produced by casein when it is adsorbed to a liquid interface \cite{dickinson:1997a,makievski:1998,maldonado:2004, casanova:2000}. Third, structurally different proteins like lysozyme, $\kappa$-casein and $\beta$-casein showed a similar behavior of the coalescence time as a function of the particle radius (for a planar interface aged during 20 minutes) \cite{dickinson:1988}. Therefore the exact analytical form of the steric potential does not appear to be very significant.  Fourth, according to the experimental methodology, bare drops of hexadecane were formed by vigorous mixing with a buffer solution. Then the drops were injected into the cell containing the protein solution and the planar interface. All the drops reached the interface within the first five minutes, and frequently before $30\,$s. However, it takes hours for $\beta$-casein to attain an equilibrium concentration in the presence of a hexadecane/water interface (see Fig. 9 in Ref. \cite{dickinson:1988}). Consequently, the amount of protein adsorbed at the interface of the drops is uncertain, and the exact magnitude of the steric barrier is unknown. As a result of all these factors, it was considered convenient to simplify the form of the potential as much as possible, and test the effect of different steric barriers on the outcome of the simulations. 
 
\begin{table*}[th!]
\centering
\begin{tabular}{c|l}
\hline\hline
\raisebox{1.0ex}{}\\
&\raisebox{-1.0ex}{$V_\mathrm{vdW}=-\frac{A_H}{12}\left[\frac{y}{x^2+xy+x}+\frac{y}{x^2+xy+x+y}+2 \ln\left(\frac{x^2+xy+x}{x^2+xy+x+y}\right)\right].$ \hspace{2.88cm} \cite{hamaker:1937}}\\
&\raisebox{-3.0ex}{$V_\mathrm{elect}=\frac{64\pi}{\kappa}C_\mathrm{el}kT\tanh\left(\frac{e\Psi_\mathrm{si}}{4kT}\right)\tanh\left(\frac{e\Psi_\mathrm{sj}}{4kT}\right)\times e^{-kh}\left[\frac{2r_ir_j}{\kappa(r_i+r_j)}\right].$ \hspace{2.75cm}\cite{danov:1993a}}\\
\raisebox{-0.0ex}{Spherical}
&\raisebox{-3.0ex}{$V_\mathrm{st}=\frac{4kT}{3V_1}\bar{\phi}_i\bar{\phi}_j\left(\frac{1}{2}-\chi\right)\left(\delta-\frac{h}{2}\right)^2\left[\frac{3(r_i+r_j)}{2}+2\delta+\frac{h}{2}-\frac{3(r_j-r_i)^2}{2(h+r_i+r_j)}\right],$ \hspace{2.15cm}\cite{lozsan:2005,lozsan:2006}}\\
&\raisebox{-1.0ex}{\hspace{9cm} $\delta < h < 2\delta$.}\\
&\raisebox{-3.0ex}{$V_\mathrm{st}=\frac{kT}{V_1}\left(\frac{1}{2}-\chi\right)\left[\left(\bar{\phi}_j\right)^2\left(\frac{v_a^2}{v_c}-v_a\right)+\left(\bar{\phi}_i\right)^2\left(\frac{v_b^2}{v_c}-v_b\right)+\,2\bar{\phi}_i\bar{\phi}_j\left(\frac{v_av_b}{v_c}\right)\right],$ \hspace{1.25cm}\cite{lozsan:2006}}\\
&\raisebox{-1.0ex}{\hspace{9cm} $0 < h < \delta$.}\\
\raisebox{1.0ex}{}\\
\hline\hline 
&\raisebox{-4.0ex}{$V_\mathrm{vdW}=-\frac{A_H}{12}\left\{\frac{2r_j(l-h)}{l(L+h)}+\frac{2r_j(l-h)}{h(l+L)}\right.+2\ln\left[\frac{h(l+L)}{l(h+L)}\right]+\frac{r_\mathrm{f}^2}{h^2}$} \\
&$\hspace{1.1cm} -\frac{l-h}{L}\frac{2r_\mathrm{f}^2}{hl}-\frac{l-r_i-(L-r_j)}{2l-2r_i-h}\frac{2r_\mathrm{f}^2}{hl}-\frac{2(L-r_j)-h}{2l-2r_i}\frac{d-h}{2h}+\frac{2r_jL^2(l-h)}{hl(l+L)(L+h)}$\\
&$ \hspace{1.1cm} -\frac{2r_j^2}{h(2l-2r_i-h)}\frac{l^2+r_\mathrm{f}^2}{(l+L)(l+L-2r_j)}+\frac{2r_j^2d}{(2l-2r_i-h)\left[(h+L)(h+L-2r_j)-(l-h)(l-2r_i-h)\right].}$\\
&$\hspace{1.1cm}\left.-\frac{4r_j^3(l-h)}{(l+L)(l+L-2r_j)\left[(h+L)(h+L-2r_j)-(l-h)(l-2r_i-h)\right]}\right\}.$ \hspace{3.45cm} \cite{danov:1993a}\\
\\
&\raisebox{-1.0ex}{$V_\mathrm{elect}=\frac{64\pi}{\kappa}C_\mathrm{el}kT\tanh\left(\frac{e\Psi_\mathrm{si}}{4kT}\right)\tanh\left(\frac{e\Psi_\mathrm{sj}}{4kT}\right)\times e^{-kh}\left[r_\mathrm{f}^2+\frac{2r_ir_j}{\kappa(r_i+r_j)}\right].$ \hspace{2.2cm}\cite{danov:1993a}}\\
\raisebox{-3.0ex}{Deformable}
&\raisebox{-3.0ex}{$V_\mathrm{st}=\frac{4kT}{3V_1}\bar{\phi}_i\bar{\phi}_j\left(\frac{1}{2}-\chi\right)\left(\delta-\frac{h}{2}\right)^2\left[\frac{3(r_i+r_j)}{2}+2\delta+\frac{h}{2}-\frac{3(r_j-r_i)^2}{2(h+r_i+r_j)}\right],$ \hspace{2.27cm}\cite{lozsan:2006}}\\
&\raisebox{-1.0ex}{\hspace{9cm} $\delta < h < 2\delta$.}\\
&\raisebox{-3.0ex}{$V_\mathrm{st}=\frac{kT}{V_\mathrm{w}}\left(\frac{1}{2}-\chi\right)\left[\left(\bar{\phi}_j\right)^2\left(\frac{v_a^2}{v_c}-v_a\right)+\left(\bar{\phi}_i\right)^2\left(\frac{v_b^2}{v_c}-v_b\right)+\,2\bar{\phi}_i\bar{\phi}_j\left(\frac{v_av_b}{v_c}\right)\right],$ \hspace{1.32cm}\cite{lozsan:2006}}\\
&\raisebox{-1.0ex}{\hspace{9cm} $0 < h < \delta$.}\\
&\raisebox{-2.0ex}{$V_\mathrm{dil}=\frac{\pi\gamma_0r_\mathrm{f}^4}{2r_a^2}$. \hspace{9.5cm} \cite{danov:1993a}}\\
\\
&\raisebox{-2.0ex}{$V_\mathrm{bend}=-\frac{2\pi B_0r_\mathrm{f}^2}{r_a}, \, (r_\mathrm{f}/r_a)^2\ll 1$. \hspace{6.8cm} \cite{ivanov:1999}}\\
\raisebox{1.0ex}{}\\
\hline\hline
\end{tabular}
\caption{Potentials used in the simulations. In these equations, $r_i$ is the radius of the small droplet, $r_j$ is the radius of the large drop and $h$ is the minimum distance between their surfaces.  For the van der Waals (vdW) potential: $A_H$ is the Hamaker constant, $x=\frac{h}{2r_i}$, $y=\frac{r_i}{r_j}$, $l=h+r_i+\sqrt{r_i^2-r_\mathrm{f}^2}$, $L=r_j+\sqrt{r_j^2-r_\mathrm{f}^2}$, $d=\sqrt{h^2+4 r_\mathrm{f}^2}$, $h$ and $r_\mathrm{f}$ are the thickness and radius of the film, respectively. For the electrostatic potential (elect): $\kappa^2=\frac{8\pi e^2 z^2}{\epsilon kT}C_\mathrm{el}$, $z$ is the charge number, $\epsilon$ is the dielectric permittivity of the medium, $C_\mathrm{el}$ is the electrolyte concentration, $e$ is the electron charge, $kT$ is the thermal energy, $\Psi_\mathrm{si}$ and $\Psi_\mathrm{sj}$ are the surface potentials for the small and large drops, respectively. For the steric potential (st): $V_\mathrm{w}$ is the molar volume of the solvent, $\chi$ is the Flory-Huggins solvency parameter of the protein, $\bar{\phi}_j$ and $\bar{\phi}_i$ are the average volume fraction of the protein around each sphere, $\bar{\phi}_i=\frac{3r_i^2\Gamma M_\mathrm{p}}{\rho_\mathrm{p} N_A \left[(r_i+\delta)^3-r_i^3\right]}$, with $\Gamma$ the number of molecules per unit area, $\rho_\mathrm{p}$ the density of the protein, $M_\mathrm{p}$ the molecular weight of the protein and $\delta$ the width of the protein layer. Volumes $v_a$, $v_b$, and $v_c$ depend on $h$. Their explicit geometrical expressions to calculate the volume of overlap between the interacting particles can be seen in Ref. \cite{lozsan:2006}. For the dilatational (extensional) potential (dil):  $\gamma_0$ is the interfacial tension  and $r_a=\frac{2r_i r_j}{r_i+r_j}$. For the bending potential (bend): $B_0=1.6\times 10^{-12}\,$N \cite{kralchevsky:1991a, kralchevsky:1991b} is the interfacial bending moment.}
\label{potentials}
\end{table*}

\begin{figure}[th!]
\subfigure[]{
\includegraphics[scale=0.3]{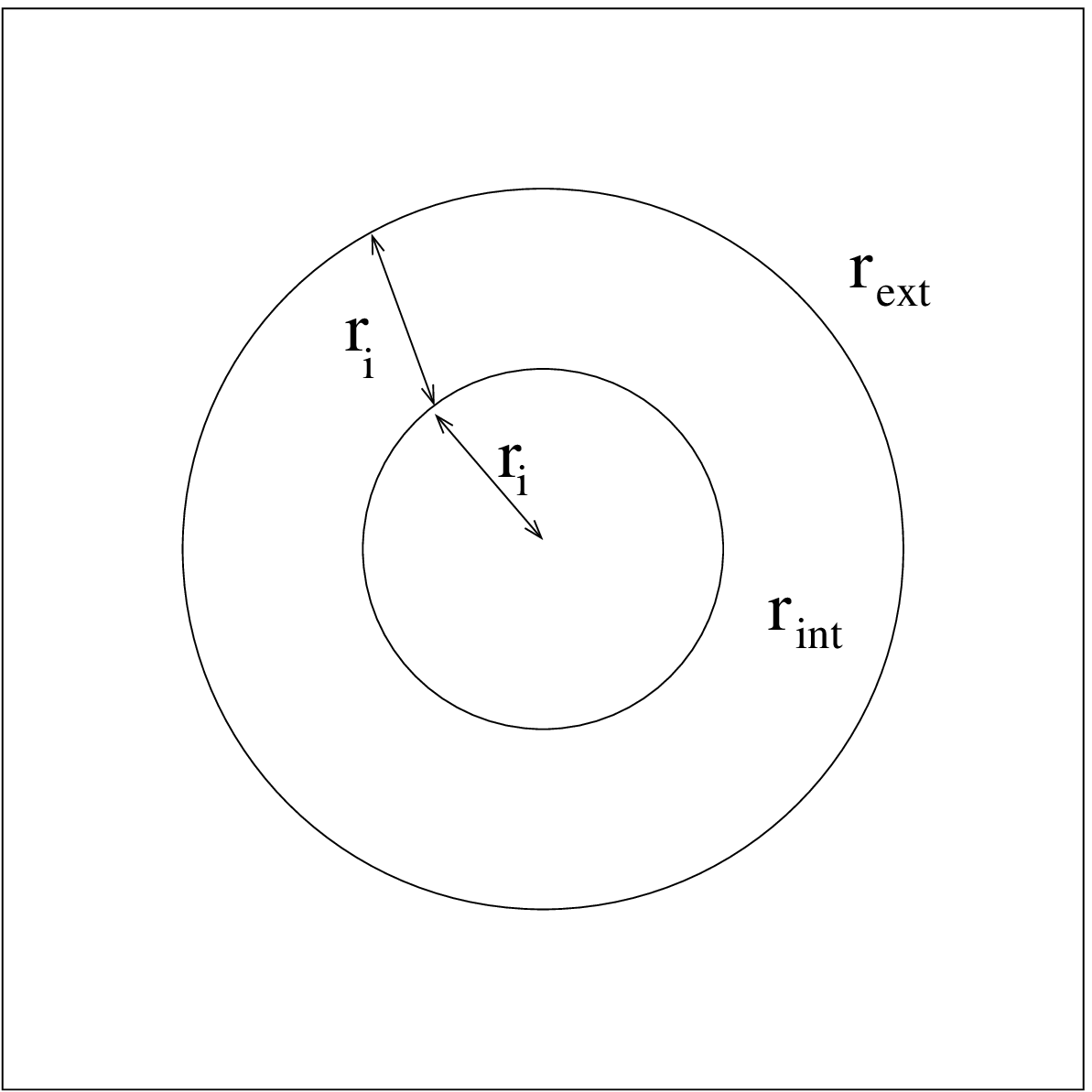}}
\subfigure[]{
\includegraphics[scale=0.3]{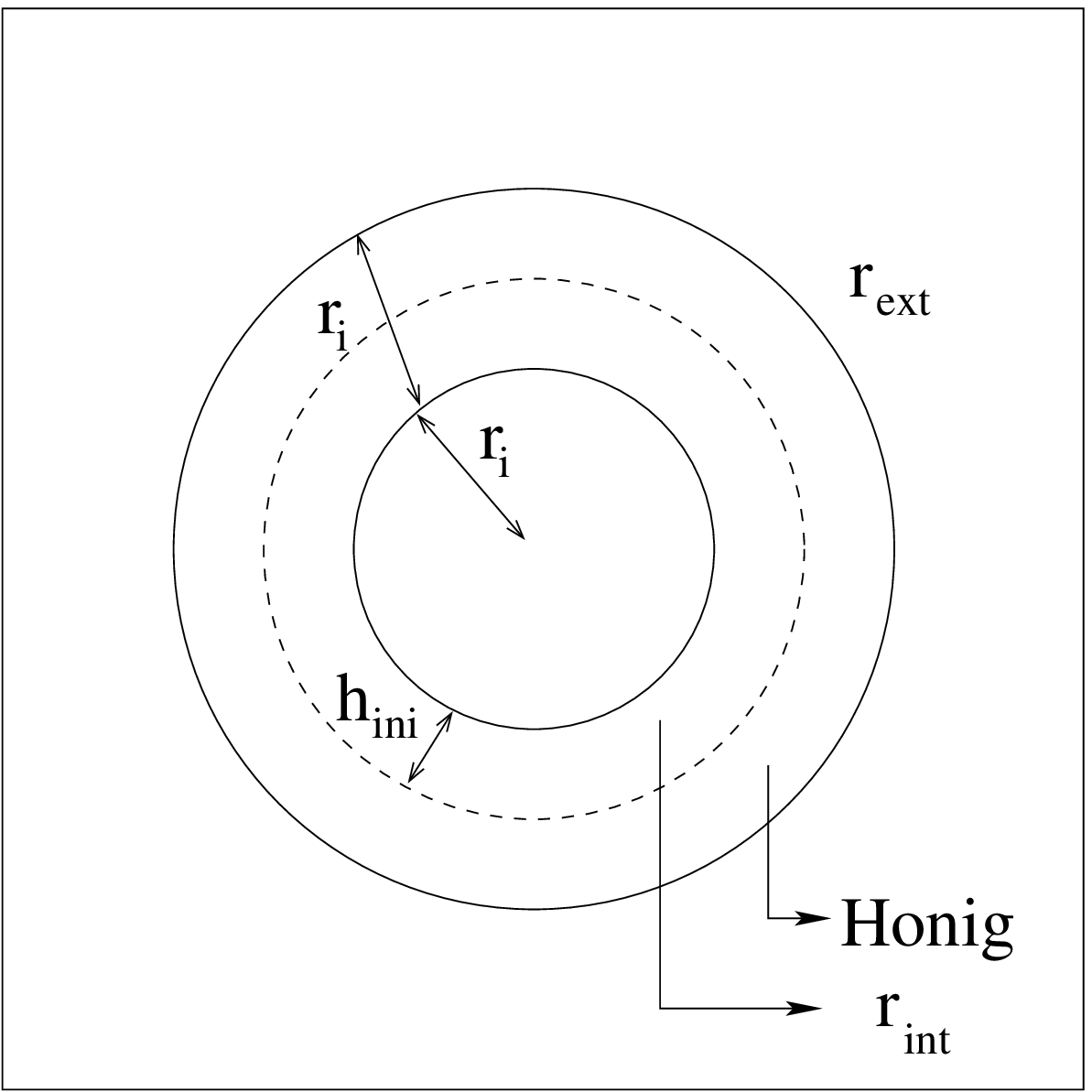}}\\
\caption{Calculation of the diffusion tensor. (a) Spherical: Inner region ($0<h\leq r_i$), outermost region ($h>r_i$),  $h=r_{ij}-r_i-r_j$.  (b) Deformable: Inner region ($0<h \leq h_\mathrm{ini}$), intermediate region  ($h_\mathrm{ini}<h\leq r_i$), outermost region ($h>r_i$).}
\label{ccfloc}
\end{figure}

Fig. \ref{ccfloc} illustrates the methodology employed for evaluating the diffusion tensor of the drops. In the case of spherical drops the space around each drop $i$ is divided into two regions. An imaginary sphere of radius $d_\mathrm{int}= 2r_i$ delimits the internal region. If a neighbor particle reaches the internal region of particle $i$ (case I): $h<r_i$ (with $h=r_{ij}-r_i-r_j$), the position of the closest particle is used to calculate the diffusion constant of $i$. The program has the possibility to select between the expressions of Honig \cite{honig:1971}, Taylor \cite{taylor:1923}, and a linear combination of the expressions of Taylor for immobile and mobile interfaces. If none of the surrounding particles reaches the internal region of particle $i$ (case $II$), the expressions of Stokes, Mills and Snabre \cite{mills:1994}, Honig {\it et al.} \cite{honig:1971}, Taylor \cite{taylor:1923}, or Beenakker {\it et al.}  \cite{beenakker:1982,beenakker:1984,urbina:2003}, can be assigned to particle $i$. The formulas of Mills {\it et al.} and Beenakker {\it et al.} use the volume fraction of oil in the simulation box to evaluate an empirical form of the diffusion tensor. 
 
In the case of deformable drops the options of Stokes, Mills {\it et al.}, Beenakker {\it et al.}, and Honig {\it et al.} are also available if all the neighbor particles are located within the external region of drop $i$. However, the internal region around each particle is now subdivided into two parts. The outermost zone of the internal region corresponds to Region I defined above in terms of the approximation distance between two deformable drops. In this region the deformable drops are close to one another but still maintain their spherical shape, $h_\mathrm{ini}<h\leq r_i$. Consequently the expression of Honig {\it et al.} is used to evaluate the diffusion tensor. The innermost zone of the internal region of particle $i$ corresponds to the regions of approach II and III defined above. Within these regions, the drops behave as truncated spheres. Hence, the set of formulas compiled by Gurkov and Basheva for deformable drops (see Table 1 in Ref. \cite{gurkov:2002}), can be selected. 

It might occur that the range of validity of the mathematical expression used to simulate the diffusion tensor at close separation distances is exceeded in some cases. The program checks if either $f_\c^{(1)}>1$ or $f_\c^{(2)}>1$. In that case, it has the option to correct the diffusion tensor using the expression of Honig {\it et al.}, or use the calculated diffusion constant regardless of its value.

The analytical form of the set of tensors used in the present simulations are shown in Table \ref{diffusion} (see the section on Computational Details).

\begin{table*}[htbp]
\centering
\begin{tabular}{c|l|l}
\hline\hline
&Geometry & Tensor\\
\hline\hline
\raisebox{-2.0ex}{$I$}&
\raisebox{-2.0ex}
{Stokes Immobile: A sphere in an}
&\raisebox{-2.0ex}{$D_\mathrm{St}=D_0$.}\\ 
&unbounded liquid \cite{basheva:1999}.\\
\raisebox{-2.0ex}{$II$}&
\raisebox{-2.0ex}
{Honig \cite{urbina:2009b}.}
&\raisebox{-2.0ex}{$D_\mathrm{Hn}=\frac{6u^2+4u}{6u^2+13u+2}D_0$.} \\
\raisebox{-2.0ex}{$III$}&
\raisebox{-2.0ex}
{Taylor Immobile: Two spheres of radii}
&\raisebox{-2.0ex}{$D_\mathrm{Ta}=4D_0\frac{r_i}{r_*}h$.}\\
&$r_i$,$r_j$, and $h\ll r_i,r_j$ \cite{gurkov:2002}.\\
\raisebox{-2.0ex}{$IV$}&
\raisebox{-2.0ex}
{Taylor Mobile:  Two sphererical liquid}
&\raisebox{-2.0ex}{$D_\mathrm{Ta,m}=4D_0\frac{r_i}{r_*}h\left(\frac{1-1.711\xi-0.461\xi^2}{1-0.402\xi}\right)$}.\\
&drops or radii $r_i$, $r_j$ \cite{gurkov:2002}.\\
\raisebox{-2.0ex}{$V$}&
\raisebox{-2.0ex}
{Reynolds Immobile: Film between two}
&\raisebox{-2.0ex}{$D_\mathrm{Re}=4r_iD_0\frac{h^3}{r_\mathrm{f}^4}$.}\\
&circular disks or  radius $r_\mathrm{f}$ and $h\ll r_\mathrm{f}$ \cite{gurkov:2002}.\\
\raisebox{-2.0ex}{$VI$}&
\raisebox{-2.0ex}
{Two deformed drops \cite{danov:1993b}.}
&\raisebox{-2.0ex}{$D_\mathrm{Dd}=\frac{4h}{r_i}\left(1+\frac{r_\mathrm{f}^2}{r_i h}+\frac{\epsilon_\mathrm{S} r_\mathrm{f}^4}{r_i^2 h^2}\right)D_0$.}
\raisebox{-0.5ex}{}\\
\hline\hline
\end{tabular}
\caption{Tensors used in the simulation. Here $r_i$ is the radius of the small droplet, $r_j$ is the radius of the large drop, $r_*=\frac{2r_ir_j}{r_i+r_j}$, $h$ and $r_\mathrm{f}$ are the thickness and radius of the film, $D_0=\frac{kT}{6\pi\eta r_i}$, $\xi=\frac{\eta}{\eta_i}\sqrt{\frac{r_i}{h}}$, $\eta$ is the dynamic viscosity of the continuous phase, $\eta_i$ is the dynamic viscosity of the disperse phase, $u=\frac{r_{ij}-r_i-r_j}{R_0}$, where $r_{ij}=r_i+r_j+d$  is the distance between centers with $R_0$ as a radius of reference and $\epsilon_\mathrm{S}$ have values between $0.001$ and $1$.}
\label{diffusion}
\end{table*}

While the coalescence of non-deformable drops occurs when the particles overlap, deformable drops could coalesce through two different mechanisms: 

\begin{enumerate}
\item The drainage of the intervening film between flocculated drops. In this case the drops move until the distance of separation between their surfaces reaches $h=h_\mathrm{crit}$.

\item Surface oscillations and/or the formation of holes promote the rupture of the film prior to its drainage \cite{vrij:1964,vrij:1966,scheludko:1967,vrij:1968,kashchiev:1980,velikov:1997, nikolova:1999,manev:2005a}. In this case, the surface oscillations of the film are not simulated explicitly, but a stochastic procedure is implemented, taking into account some relevant aspects of the rupture process. 

First, the lifetime of a doublet, ($\tau_{ij}$), is calculated using one out of two procedures:

\begin{description}
\item{2.1.} Counting the time continuously from the moment a doublet enters regions II or III until either $h=h_\mathrm{crit}$ (coalescence occurs) or the doublet separates: $r_{ij}>r_i+r_j+h_\mathrm{ini}$. 

\item{2.2.} Adding the time steps in which each couple of particles enter the regions II or III. In this case, $\tau_{ij}$ is different from zero after a pair of particles enter regions II or III the first time. This way of counting the time assumes that the probability of rupture is proportional to the time spent by the doublet in the region of deformation.

\end{description}

Following, a random number between $-1.0$ and $1.0$ is assigned to the surface of each drop. The amplitude of each capillary wave ($A_i$) is estimated as the product of the referred random number times the value of $h_\mathrm{crit}$:

\begin{equation}
A=\mathrm{ran(t)}\times h_\mathrm{crit}.
\end{equation}

If the mechanism of surface oscillations is activated, the value of $\tau_{ij}$ is compared at each time step with a characteristic time deduced by Vrij \cite{vrij:1968} for the fastest increase of surface oscillations: 

\begin{equation}
\tau_\mathrm{Vrij}=96\pi^2\gamma\eta h_\mathrm{ini}^5 A_H^{-2}.
 \label{vrij}
\end{equation}

Eq. (\ref{vrij}) was deduced assuming van der Waals interactions only. A more general expression for $\tau_\mathrm{Vrij}$ \cite{vrij:1968} requires knowledge of the second order differential of the free energy in terms of $h$, under certain mathematical restrictions. Hence, it is very difficult to calculate when the model of deformable drops is employed. Therefore the program has the option to introduce the value of $\tau_\mathrm{Vrij}$ as part of the input data. 

The value of the tension in Eq. (\ref{vrij}) is approximated by the average of the interfacial tension of the two drops ($\gamma_i+\gamma_j$)/$2$. The value of $\gamma_i$ is then calculated at each time step of the simulation, using the number of surfactant molecules adsorbed:

\begin{equation}
 \gamma_i=\gamma_0+(\gamma_\mathrm{cmc}-\gamma_0)\left(\frac{N_{\mathrm{s},i}}{N_{\mathrm{s},i}^\mathrm{max}}\right).
\end{equation}

Here $\gamma_0$ and $\gamma_\mathrm{cmc}$ stand for the value of the O/W interfacial tension in the absence of surfactant molecules, and at the CMC of the surfactant employed. $N_{\mathrm{s},i}$ being the number of surfactant molecules adsorbed to the interface of drop $i$, and $N_{\mathrm{s},i}^\mathrm{max}$, the maximum number of surfactants that can be adsorbed to that drop.  When the surfactant concentration is equal to zero, $N_{\mathrm{s},i}$=0 at all times, and $\gamma_i=\gamma_0$. In the case in which the surfactant concentration is enough to cover the drops completely, the equilibrium surface tension $\gamma_i=\gamma_\mathrm{cmc}$ is reached.
   
Coalescence occurs whenever the total height of the surface oscillations is greater than $h_\mathrm{crit}$. The height of the oscillations is approximated by:

\begin{equation}
A_\mathrm{TOT}=(A_i+A_j)\left[\exp\left(\frac{\tau_{ij}}{\tau_{Vrij}}\right)-1\right].
 \label{lambda}
\end{equation}

Eq. (\ref{lambda}) takes into account that: (a) the surface oscillations increase exponentially with time \cite{vrij:1968,vrij:1964,vrij:1966}, and (b) the capillary waves can be in-phase or out-of-phase. 

\end{enumerate}

\section{Computational Details}
\label{computational}

\begin{figure}[thbp]
\centering
\includegraphics[scale=0.25]{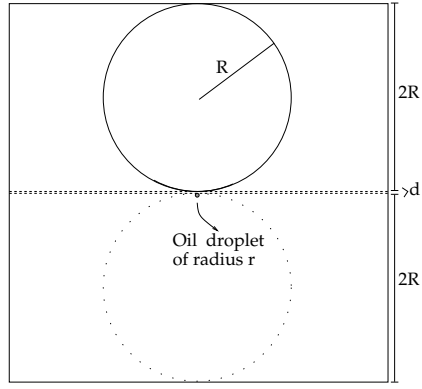}
\caption{Model of the drop/interface system employed in the simulations. Here, $r$ is the radius of the small droplet, and $R$ the radius of the large drop resembling the interface. $d$ is the initial distance between the surfaces of the drops.}
\label{model}
\end{figure}

In order to reproduce the lifetime of a drop of hexadecane at the water/hexadecane interface, two drops of very different sizes are used. The radius of the small droplet $r$ ($r_i$ = $r$), was varied between $1$ to $10\,\mu$m.  A fixed drop of $500\,\mu$m represents the interface. A square simulation box with a side length $L$ equal to $4\,R+d$ (where $d$ is the initial distance between the drops, and $R$ ($r_j$ = $R$) is the radius of the large drop) was employed. Figure \ref{model} illustrates the proportion between the big drop and the moving droplet, as well as the spatial position of the doublet at the beginning simulation. The mean coalescence time and its standard deviation was estimated using $1000$ random walks for each particle size.  For each individual simulation the initial values of the random term in Eq. (\ref{brownian}), $\mathbf{R}$, were changed in order to favor different trajectories. The logistic equation in the region of chaos was used to avoid the saturation of the random number generator (see Ref. \cite{urbina:2009b} for details). 

The surfactant concentration of $\beta$-casein and the ionic strength of the aqueous solution employed in the simulations correspond to the ones used in the experimental measurements of Dickinson {\it et al.} ($10^{-4}$ wt$\%$ protein, pH=$7.0$, ionic strength $0.1\,$M, T=$298\,$K \cite{dickinson:1988}). 

As pointed out by Damodaran and Rao \cite{damodaran:2001}, the interfacial areas of the proteins resulting from $\Gamma$ {\it vs.} $t$ or $\Pi$ {\it vs.} $t$ data are considerably smaller than the cross sectional areas of the native or unfolded proteins, which typically fall in the range $100-1000\,$nm$^2$ \cite{damodaran:2001}.  According to the experimental data of Graham and Phillips \cite{graham:1979a,graham:1979b,graham:1979c}, the maximum surface coverage of $\beta$-casein at the air/water and the oil/water interface occurs between $2-3\,$mg/m$^2$. Chen and Dickinson, determined the value of $\Gamma$ for a hexadecane-in-water emulsion stabilized with $0.4$\% wt $\beta$-casein, obtaining a value of $2.95\,$mg/m$^2$ \cite{chen:1993a,chen:1993b}. This value suggests an interfacial area of the order of $14\,$nm$^2$.  Conversely, one can estimate the interfacial area of a $\beta$-casein micelle from its radius ($13 \,$nm \cite{dimitrova:2004,leclerc:1997}), and compute the number of molecules in the micelle from its molecular weight ($250\,$kDa \cite{dimitrova:2004,dickinson:1997b}). In this case, a much larger interfacial area of $204\,$nm$^2$ is obtained. This discrepancy can be partially due to the internal structure of the casein micelles \cite{tuinier:2002,kruif:1998}, and the large number of conformations available for a long polyelectrolyte when it is partially dissolved in a liquid medium.   

The analytical form of tensors employed in the present simulations, $D_\c$, are shown in the Table \ref{diffusion}. Following the scheme of Fig. \ref{ccfloc}, two tensors must be defined for spherical and deformable droplets, respectively. The first one corresponds to the outer region of approach denoted by $r_\mathrm{ext}$ in Fig.  \ref{ccfloc}. The second one corresponds to the inner region of approach ($r_\mathrm{int}$ in Fig. \ref{ccfloc}). 

In the case of spherical droplets the following combinations of tensors $D_\mathrm{ext}/D_\mathrm{int}$, were tested:

\begin{description}
\item{A.} Stokes/Honig. Tensors $I$ and $II$ of Table \ref{diffusion} are used. The radius of reference $R_0$ for the equation of Honig {\it et al.} was approximated by the average radius of the drops: $R_0=(r_i + r_j)/2$.
\item{B.} Honig/Honig. Tensors $I$ and $II$ of Table \ref{diffusion} are used with $R_0=(r_i + r_j)/2$.
\item{C.} Honig/Honig. Tensor $II$ of Table \ref{diffusion} is used in both the internal and external regions of drop $i$, with $R_0=r_i$, where $r_i$ is the radius of the smallest drop.
\item{D.} Stokes/Taylor Immobile. Tensors $I$ and $III$ of Table \ref{diffusion} are used. 
\item{E.} Stokes-Taylor Immobile/Taylor Immobile. Tensors $I$ and $III$ of Table \ref{diffusion} are used. The tensor of Taylor is used in the external region as long as $f_\c^{(2)}=D_\t/D_0 \le 1$. Otherwise, the expression of Stokes is used.
\item{F.} Taylor Immobile/Taylor Immobile. Tensor $III$ of Table \ref{diffusion} is used regardless of the value of $f_\c^{(2)}$. 
\item{G.} Stokes/Taylor Mobile. Tensors $I$ and $IV$ of Table \ref{diffusion} are employed. Values of $3.032\times 10^{-3}\,$Pas and $8.905\times 10^{-4}\,$Pas were used for the viscosity of hexadecane \cite{wang:2004,nhaesi:2000} and water at $T = 298.15\,$K, respectively.
\item{H.} Stokes/Taylor Mobile-Taylor Immobile. Tensor $I$ of Table \ref{diffusion} is used for the external region. Within the internal region a linear combination of tensors $III$ and $IV$ are used:

\begin{equation}
D_i=C_\mathrm{TR} D_\t+(1.0-C_\mathrm{TR})D_{\t,m},
\label{mobimm}
\end{equation}
where $C_\mathrm{TR}=N_{\mathrm{s},i}/N_{\mathrm{s},i}^\mathrm{max}$. It is clear from Eq. (\ref{mobimm}), that the total coverage of a drop generates the use of the Taylor's formula for immobile interfaces. However, it is known that the adsorption of $\beta$-casein to an hexadecane/water interface is very slow \cite{dickinson:1988}. Hence, it is likely that in the experiments of Dickinson {\it et al.}, the protein layer at the surface of the emerging drop is very dilute even at the moment of the collision with the planar interface. This favors an intermediate mobility of liquid at its interface, similar to the one described by Eq. (\ref{mobimm}). 
In the present simulations the temporal dependence of the protein adsorption was not considered. Hence, an intermediate interfacial mobility Eq. (\ref{mobimm}) can only be produced if either the amount of protein molecules dissolved in the aqueous phase is subtracted from the total surfactant population, or if a lower protein concentration is arbitrarily introduced in the simulations. Consequently, in order to study the effect of tensor H on the coalescence time of hexadecane droplets, a protein concentration of $C_\mathrm{s}=10^{-9}\,$M was used. This concentration generated a value of $C_\mathrm{TR}=0.33$.
\end{description}

In the case of deformable droplets, the inner and outer regions of approach are separated by an intermediate zone. In this zone the approaching drops are still spherical. As shown in Fig. \ref{ccfloc}, the program always uses the expression of Honig {\it et al.} in this intermediate region.  
Notice also that the expression of Reynolds (Eq. $V$ in Table \ref{diffusion}) requires $h\ll r_i,r_j$. However, there might be cases in which the actual situation falls out of the range of validity of Eq. $V$. This might occur if the initial deformation distance is introduced as part of the input, or when the motion of particles of nanometer size is simulated. In order to account for this deficiency, it is assumed that the formation of a plane parallel film will necessarily delay the coalescence process. Hence, the code uses Eq. $V$ as long as $D_\mathrm{Re}/D_0 \le 1$. Otherwise, the expression of Reynolds is substituted by the one of Honig {\it et al.} in the internal region. 

The following combinations of $D_\mathrm{ext}/D_\mathrm{int}$ were tested for the case of deformable droplets:

\begin{description}
\item{I.} Stokes/Reynolds. Tensors $I$ and $V$ of Table \ref{diffusion} are used. The value of $h_\mathrm{ini}$ is calculated by the program using Eq. (\ref{def4}).
\vspace{-0.05cm}
\item{J.} Honig/Reynolds. Tensors $II$ and $V$ are used. The value of $h_\mathrm{ini}$ is calculated by the program using Eq. (\ref{def4}).
\vspace{-0.05cm}
\item{K.} Honig/Reynolds. Tensors $II$ and $V$ are used. The value of $h_\mathrm{ini}$ is set equal to $20\,\mu$m and the value of  $r_\mathrm{f}=r_i/10$.
\vspace{-0.05cm}
\item{L.} Honig/Reynolds. Tensor $II$ and $V$ are used. The value of $h_\mathrm{ini}$ is set equal to $20\,\mu$m and the value of  $r_\mathrm{f}=r/100$.
\vspace{-0.05cm}
\item{M.} Stokes/Danov. Tensors $I$ and $VI$ are employed. The value of $h_\mathrm{ini}$ was calculated by the program using Eq. (\ref{def4}). Values for $\epsilon_\mathrm{S}=1$ and $0.1$ were tested.
\end{description}

For every combination of tensors a recursive calculation of the coalescence time was executed $1000$ times. The results were compared with the experimental coalescence times obtained by Dickinson {\it et al.} These times were obtained by digitalization of Fig. 4 of Ref. \cite{dickinson:1988} using the {\it Engauge Digitizer}. This data is represented by stars in Figs. \ref{spherical}-\ref{final}. 

Most of the parameters employed in the simulations are shown in Table \ref{parameters}. The mass of a casein protein was taken from M\"obius: $24\,$kDa \cite{mobius:1998}. The interfacial area was calculated using the radius ($13$ $\,$nm) and weight ($250\,$kDa) of a spherical $\beta$-casein micelle \cite{dimitrova:2004,leclerc:1997}. The effective charge of the protein was evaluated varying the surface charge of a small hexadecane drop covered with $\beta$-casein, in order to reproduce its surface potential $\zeta$=$-27.8\,$mV ($\sigma=0.002161\,$C/m$^2$) \cite{dimitrova:2004}. Notice that the effective charge of the protein shown in Table \ref{parameters}: $2.864$e-, is substantially lower than the value reported by Dickinson $15$e- \cite{dickinson:1999} for the same molecule at pH$=7.0$. In previous works, the described procedure generated an effective charge of $0.21$e- for every single formal valence of a surfactant molecule. The value obtained for casein is $13.8$ times higher than a single effective charge. This is reasonable in the light of our previous results \cite{urbina:2009a,urbina:2006}.

\begin{table}[th!]
\centering
\begin{tabular}{cc}
\hline\hline
\raisebox{-2.0ex}
{Hamaker constant \cite{israelachvili:1998}}
&\raisebox{-2.0ex}{$4.90\times 10^{-21}\,$J}\\
\raisebox{-1.0ex}
{Ionic strength \cite{dickinson:1988}}      &\raisebox{-1.0ex}{$0.1$ mol/l}\\
\raisebox{-1.0ex}
{Surf. concentration \cite{dickinson:1988}}                 &\raisebox{-1.0ex}{4.16$\times$10$^{-8}\,$mol/l}\\
\raisebox{-1.0ex}
{Molecular mass \cite{mobius:1998}}         &\raisebox{-1.0ex}{$24$ kDa}\\
\raisebox{-1.0ex}
{Specific volume \cite{mcmeekin:1949}}                          &\raisebox{-1.0ex}{$0.743\,$cm$^3$/g}\\
\raisebox{-1.0ex}
{Electric charge }                          &\raisebox{-1.0ex}{$-2.864e$}\\
\raisebox{-1.0ex}{}\\
\hline\hline
\end{tabular}
\caption{Parameters of the one $\beta$-casein molecule.}
\label{parameters}
\end{table}

\medskip
It is believed that the stability of milk in the presence of large salt concentration is related to the steric barrier produced by its casein proteins. However, that surface activity is mostly ascribed to $\kappa$-casein. This molecule contains $63$ hydrophilic aminoacids which lie on the outside of the casein micelle \cite{tuinier:2002}. In the case of $\beta$-casein only $50$ out of $209$ amino acids are hydrophilic whereas the rest are mainly hydrophobic \cite{leclerc:1997}. Depending on the protein concentration and the temperature, the core density of the $\beta$-casein micelles varies between $0.4$ and $0.9\,$g/cm$^3$, whereas the density of a globular protein is close to $1.35\,$g/cm$^3$. Moreover, the density of the outer shell of the micelles is reported to be much lower, between $0.025\,$g/cm$^3$ and $0.14\,$g/cm$^3$ \cite{leclerc:1997}. 

In general, the steric potential of a protein is difficult to parametrize. $V_\mathrm{st}$ (Table \ref{potentials}) requires the width of the steric layer ($\delta$), the value of the Flory-Huggins interaction parameter ($\chi$), and the volume fraction of protein in the steric layer. 

Dickinson {\it et al.} \cite{dickinson:1993} studied the adsorption of $\beta$-casein at air/water and oil/water interfaces using Neutron Reflectivity. According to those measurements, the distribution of protein normal to the interface is well described by a dense inner layer of $2\,$nm thickness located directly at the interface, and a more tenuous secondary layer of thickness $5$-$7\,$nm extending into the aqueous phase. Hence, we selected a conservative value of $\delta$=$6.2\,$nm for the width of the polymer layer outside the drop.   

The theory of Scheutjens-Fleer was used in a previous theoretical study regarding adsorbed $\beta$-casein \cite{dickinson:1997a}. That formalism ascribes different values of the Flory-Huggins parameters to different segments of the molecule, depending on their chemical nature. We did not find either an experimental determination or a theoretical evaluation of $\chi$ for the whole $\beta$-casein. It can be observed in Table \ref{potentials}, that the difference ($1/2-\chi$) is directly proportional to the value of the steric potential. Hence, $\chi$ has to be lower than $0.5$ in order produce a repulsive barrier ($0<\chi<0.5$). Tuinier and Kruif used the theoretical expression of Alexander-deGennes \cite{tuinier:2002} in order to account for the steric interaction between $\kappa$-casein micelles. According to that research, the height of the brush of protein is hardly affected by the solvent quality above pH $4.2$. However, the relative height of the layer with respect to its value at pH $7$, changes between $0$ and $0.2$ at pH $2$ when $\chi$ is equal to $0.5$ and $0$, respectively. A value of $\chi=0.4$ produces an intermediate relative height of $0.1$ for the brush at low pHs. In the absence of other guideline, we selected a value of $0.4$ for the Flory-Huggins parameter.

In the ESS program the value of the volume fraction of protein in the steric layer around a drop can either be introduced as an input of the simulation or calculated. As shown in Table \ref{potentials}, there is a simple equation that relates the volume fraction of protein around a drop $\bar{\phi}_i$, with the number of molecules per unit area adsorbed to its interface:

\begin{equation}
\bar{\phi}_i=\frac{3r_i^2\Gamma M_\mathrm{p}}{\rho_\mathrm{p}N_A\left[(r_i+\delta)^3-r_i^3\right]}.
\end{equation}

\begin{figure}[th!]
\centering
\bigskip
\includegraphics[angle=-90,scale=0.255]{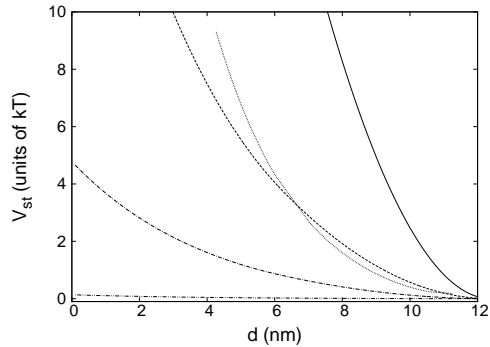}
\caption{Steric potential between two $100\,$nm $\kappa-$casein micelles.
Solid line: $V_\mathrm{st}$ ($\bar{\phi}_\mathrm{p}=2.1\times 10^{-2}$);
dashed line: $V_\mathrm{st}$ ($\bar{\phi}_\mathrm{p}=1.0\times 10^{-2})$;
dotted line: Steric potential reported by Tuinier and Kruif (Fig. 2 in Ref. \cite{tuinier:2002, kruif:1998});
dot-dashed line: $V_\mathrm{st}$ ($\bar{\phi}_\mathrm{p}=4.7\times 10^{-3}$); 
dashed double-dots line: $V_\mathrm{st}$ ($\bar{\phi}_\mathrm{p}=7.9\times 10^{-4}$).}
\label{steric}
\end{figure}

Use of the specific volume reported by McMeeking \cite{mcmeekin:1949}: $0.743\,$cm$^3$/g for a casein molecule in solution along with an interfacial area of $204\,$nm$^2$, produces the solid line illustrated in Fig. \ref{steric} for the interaction between two $\kappa$-casein micelles ($\bar{\phi}_\mathrm{p}=2.1\times 10^{-2}$). 

\begin{figure}[th!]
\centering
\bigskip
\includegraphics[angle=-90,scale=0.255]{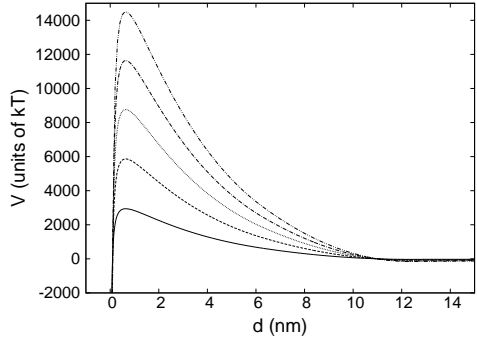}
\caption{Total potential of interaction between a large droplet of $500\,\mu$m  and a small droplet of micron-size. For this calculation the volume fraction of protein around the drops was assumed to be equal to $\bar{\phi}_\mathrm{p}=2.1\times 10^{-2}$.
Solid line: $r=2\,\mu$m;
dashed line: $r=4\,\mu$m;
dotted line: $r=6\,\mu$m;
dot-dashed line: $r=8\,\mu$m; 
dashed double-dots line: $r=10\,\mu$m.}
\label{pexp}
\end{figure}

Figure \ref{pexp} illustrates the total interaction potential between a drop of hexadecane of $10\,\mu$m and the $500\,\mu$m drop representing the planar interface. The values of $V_\mathrm{st}$ were calculated using an interfacial area per protein molecule of $204\,$nm$^2$, and a volume fraction of protein equal to $\bar{\phi}_\mathrm{p}=2.1\times 10^{-2}$. The resulting repulsive barriers are enormous. These large barriers will completely prevent the coalescence of the drop of oil with the planar interface, in contradiction with the experimental evidence. Much larger steric barriers are obtained if the interfacial area of the protein is approximated by $14\,$nm$^2$.

It is known from the experiments of Dickinson {\it et al.} \cite{dickinson:1988} that all emulsion drops coalesce with the planar interface as long as the interface is aged for only $20$ minutes. In previous papers our group demonstrated that barrier heights $\Delta V$ larger than 12.7 kT prevent the coalescence of spherical drops \cite{urbina:2009b,lozsan:2006}. The barrier heights were estimated as the difference between the values of the maximum and the secondary minimum of the total interaction potential ($\Delta V=V_\mathrm{max}-V_\mathrm{min}$).

\begin{figure}[th!]
\centering
\bigskip
\includegraphics[angle=-90,scale=0.255]{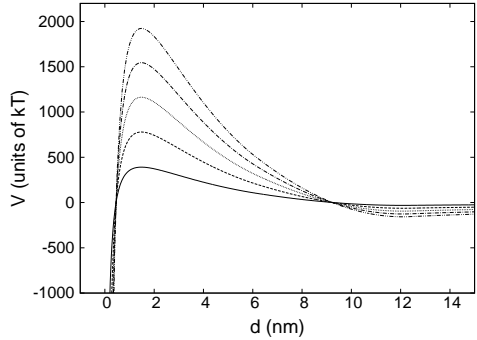}
\caption{Total potential of interaction between a large droplet of $500\,\mu$m  and a small droplet of micron-size. For this calculation the volume fraction of protein around the drops was assumed to be equal to $\bar{\phi}_\mathrm{p}=1.0\times 10^{-2}$.
 Solid line: $r=2\,\mu$m;
dashed line: $r=4\,\mu$m;
dotted line: $r=6\,\mu$m;
dot-dashed line: $r=8\,\mu$m; 
dashed double-dots line: $r=10\,\mu$m.}
\label{ppaper}
\end{figure}

It is possible to obtain surmountable repulsive barriers for the total potential varying several parameters of the steric potential. The magnitude of the electrostatic contribution is lower than the steric interaction due to the high value of the ionic strength ($I=0.1\,$M). However, the parameters of the steric potential are not independent. For instance, the volume fraction of protein around the drops depends on the width of the protein layer, which in turn depends on the surface excess of the protein at the interface ($\Gamma$).   Due to the experimental set up \cite{dickinson:1988}, the value of $\Gamma$ in the experiments of Dickinson {\it et al.} is unknown. Taking into account this uncertainty, the volume fraction of protein was systematically lowered while keeping the rest of the parameters fixed. On the one hand, it should be kept in mind that this procedure looks for an approximate form of the steric potential of casein and not for the exact value of the volume fraction. On the other hand, the calculation of the volume fraction requires knowledge of the interfacial area of the protein, its specific volume and the width of the protein layer. Hence, it summarizes the effect of several variables whose accurate values are  unknown. Moreover, the introduction of a volume fraction as an input of the simulation avoids the dependence of the calculations on the actual value of the interfacial area of the $\beta$-casein molecule (see the formula of $V_\mathrm{st}$ in Table \ref{potentials}). Notice that the electrostatic potential used in the present simulations is also independent of the interfacial area of the protein, since it was parametrized using the charge density of the drops, not the effective charge of a protein molecule. As a result, a change in the area per molecule varies the charge of a surfactant molecule but does not change the value $\sigma$, or the shape of the potential.   

\begin{figure}[th!]
\centering
\bigskip
\includegraphics[angle=-90,scale=0.255]{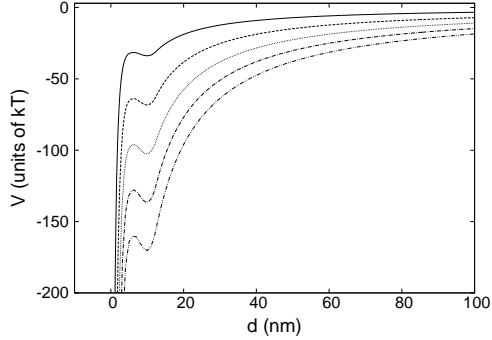}
\caption{Total potential of interaction between a large droplet of $500\,\mu$m, and a small droplet of micron-size. For this calculation the volume fraction of protein around the drops was assumed to be equal to $\bar{\phi}_\mathrm{p}=4.7\times 10^{-3}$.
 Solid line: $r=2\,\mu$m;
dashed line: $r=4\,\mu$m;
dotted line: $r=6\,\mu$m;
dot-dashed line: $r=8\,\mu$m; 
dashed double-dots line: $r=10\,\mu$m.}
\label{pop3}
\end{figure}

A volume fraction of protein of $\bar{\phi}_\mathrm{p}=1.0\times 10^{-2}$ produces an inter-micellar potential similar in magnitude to the one reported by Tuinier and Kruif (Fig. \ref{steric}). However, this volume fraction also produces large repulsive barriers between a drop and "the interface" (Fig. \ref{ppaper}).

Figure \ref{pop3} show the potentials corresponding to the system composed by a micrometer size-drop and a $500\,\mu$m drop, when a volume fraction of protein equal to $\bar{\phi}_\mathrm{p}=4.7\times 10^{-3}$ is employed. The attractive van der Waals interaction prevails. Notice that the lowest potential curve corresponds to the biggest particle size. Despite this fact, the barrier height of the potential, $\Delta V$, increases with the particle size. Barriers of $2.4\,$kT, $3.8\,$kT, $6.1\,$kT, $8.2\,$kT and $11.0\,$kT are observed for drop sizes of $r=2\,\mu$m, $r=4\,\mu$m, $r=6\,\mu$m, $r=8\,\mu$m, $r=10\,\mu$m, respectively. 

The above potential was used in some preliminary evaluations of the coalescence time. These calculations are shown in the next section. As will be shown, the small barriers shown in Figure \ref{pop3} still prevent the correct behavior of the coalescence time as a function of the particle radius.  

\section{Effect of the interaction potential, the hydrodynamic friction, and the buoyancy force}
\label{parametrization}

\begin{figure*}[th!]
\bigskip
\begin{center}
\subfigure[]{
\label{op3I}
\includegraphics[scale=0.152]{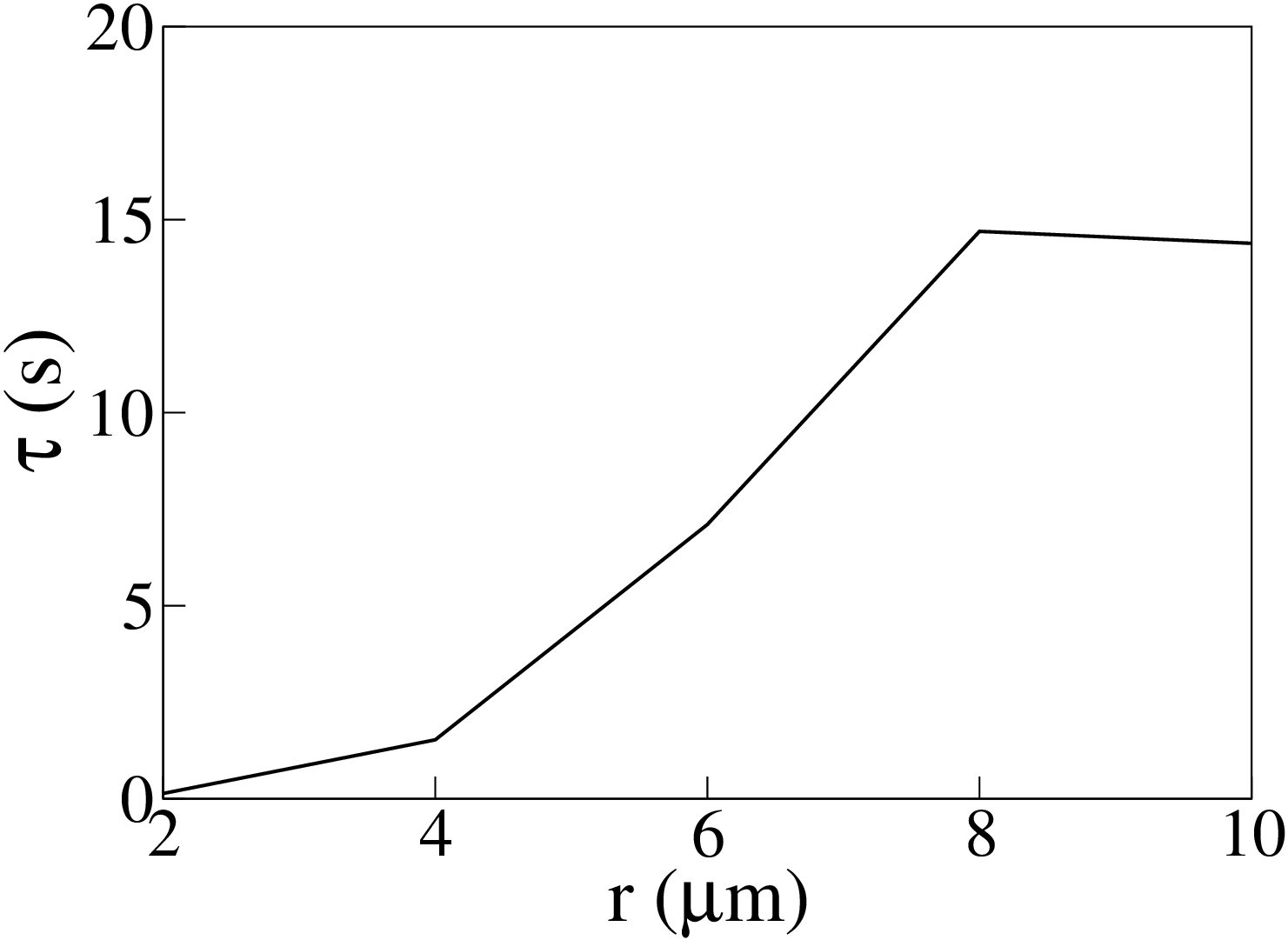}}
\subfigure[]{
\label{op3II}
\includegraphics[scale=0.152]{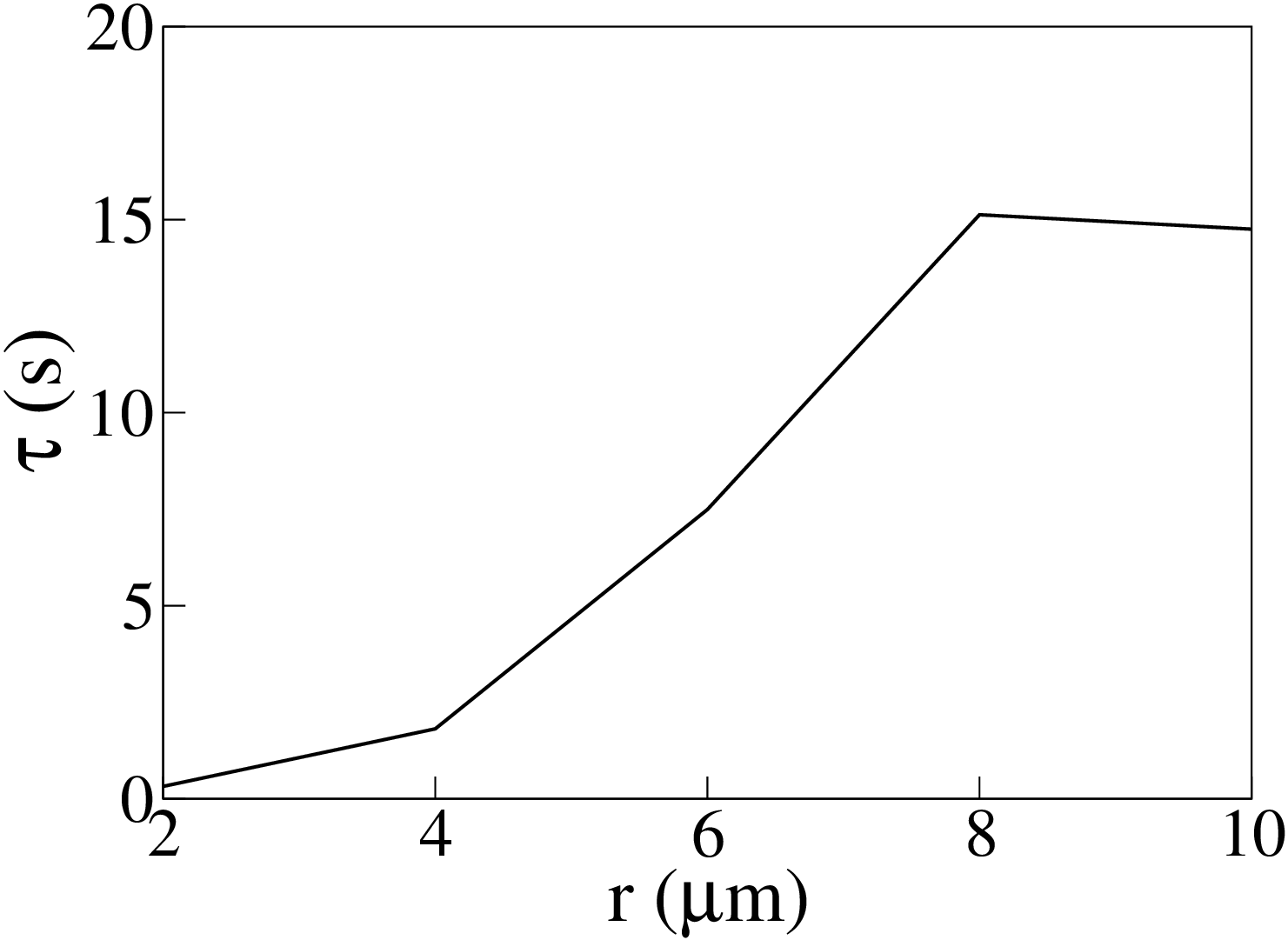}}
\subfigure[]{
\label{op3III}
\includegraphics[scale=0.152]{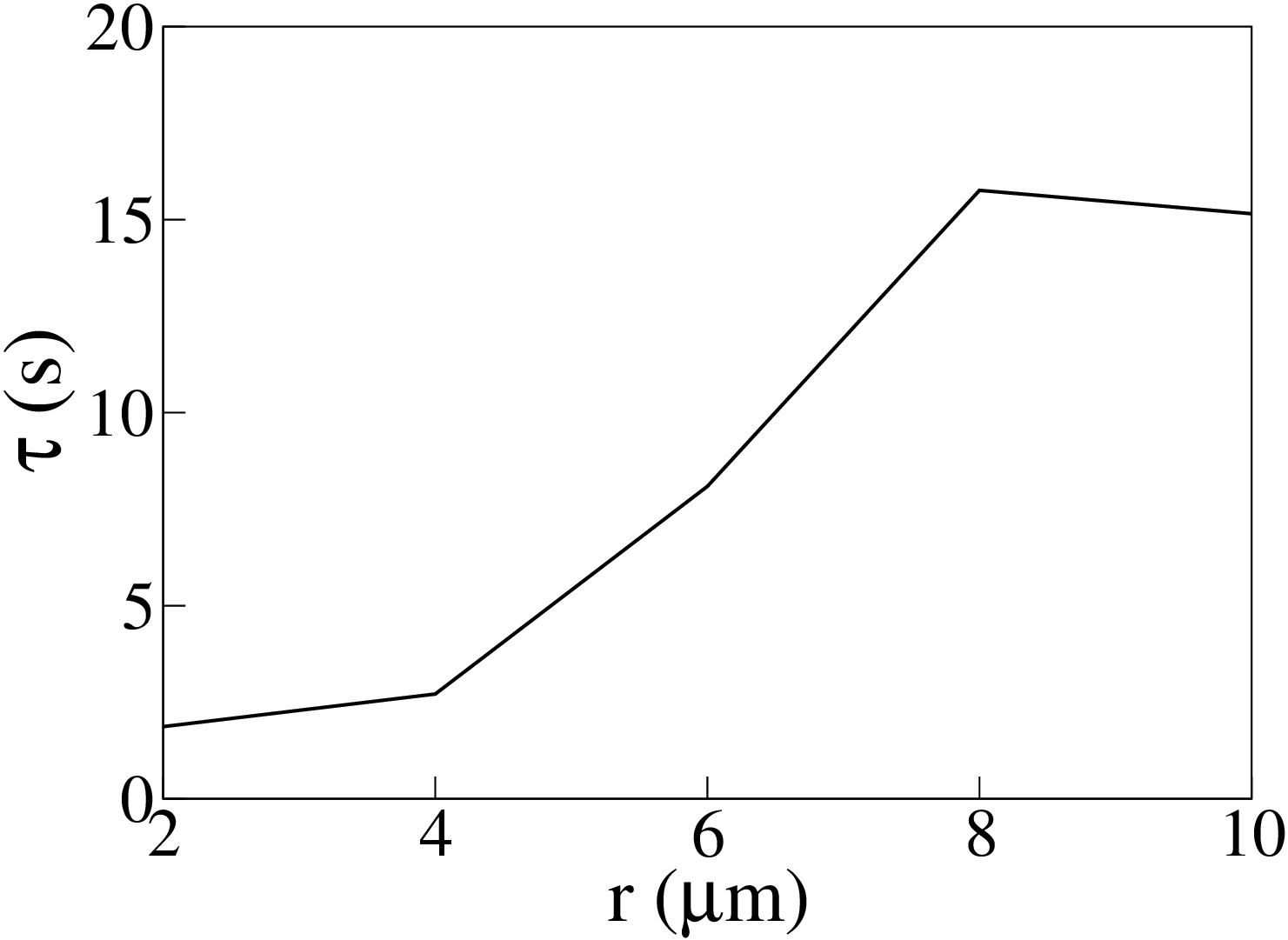}}
\bigskip
\caption{Dependence of the average coalescence time as a function of the particle radius for an initial distance of: (a) $d=20\,$nm, (b) $d=100\,$nm, (c) $d=750\,$nm. The total potential corresponds to the one depicted in Fig. \ref{pop3} (Table \ref{top3}).}
\end{center}
\end{figure*}

Table \ref{top3} shows the results of preliminary calculations in which the potentials of Fig. \ref{pop3} are employed. The average coalescence time between a drop and the interface was evaluated assuming non-deformable spherical droplets and the tensors of Stokes (outer region) and Taylor (inner zone). As shown in Fig. \ref{op3I}-\ref{op3III}, the coalescence time increases as a function of the particle radius due to the augment of the potential barrier with the particle size. 

Notice that in this case, the increase of the coalescence time shown in Figs. \ref{op3I}-\ref{op3III} is independent of the initial distance of separation between the particle and the interface ($d=20\,$nm, $100\,$nm, $750\,$nm). It is also independent of the deformability of the drops, since a similar tendency is observed for the case of deformable droplets. This trend is contrary to the effect of the buoyancy force, which promotes a decrease of the coalescence time as a function of the particle radius. However it agrees with the fact that the diffusion tensor decreases with the size of the particles, increasing their coalescence time. Thus, the increase of the repulsive barrier reinforces the effect of the hydrodynamic friction and predominates over the buoyancy force. These results are in complete contradiction with the experimental evidence that shows a decrease of the coalescence time as a function of the particle radius for micrometric drops. 

\begin{table*}[th!]
\centering 
\begin{tabular}{ccccccc}
\hline\hline
$d$ (nm) &$\tau_2$ (s) &$\tau_4$ (s) &$\tau_6$ (s) &$\tau_8$ (s) &$\tau_{10}$ (s) & Behaviour\\
\hline
 \raisebox{-1.0ex}{20}      &\raisebox{-1.0ex}{0.13751} &\raisebox{-1.0ex}{1.52714} 
&\raisebox{-1.0ex}{7.10532} &\raisebox{-1.0ex}{14.69749} &\raisebox{-1.0ex}{14.38758}
&\raisebox{-1.0ex}{Fig. \ref{op3I}}\\[1ex] 
\hline
 \raisebox{-1.0ex}{100}     &\raisebox{-1.0ex}{0.32029} &\raisebox{-1.0ex}{1.81169} 
&\raisebox{-1.0ex}{7.49218} &\raisebox{-1.0ex}{15.12774} &\raisebox{-1.0ex}{14.75408}
&\raisebox{-1.0ex}{Fig. \ref{op3II}}\\[1ex] 
\hline
 \raisebox{-1.0ex}{750}      &\raisebox{-1.0ex}{1.86863} &\raisebox{-1.0ex}{2.71276} 
&\raisebox{-1.0ex}{8.09156} &\raisebox{-1.0ex}{15.75886} &\raisebox{-1.0ex}{15.15175}
&\raisebox{-1.0ex}{Fig. \ref{op3III}}\\[1ex] 
\hline\hline
\end{tabular}
\caption{Average coalescence time for the total potential of Fig. \ref{pop3}}
\label{top3}
\end{table*}  

\begin{table*}[th!] 
\centering
\begin{tabular}{ccccccc}
\hline\hline
$d$ (nm) &$\tau_2$ (s) &$\tau_4$ (s) &$\tau_6$ (s) &$\tau_8$ (s) &$\tau_{10}$ (s) & Behaviour\\
\hline
 \raisebox{-1.0ex}{20}      &\raisebox{-1.0ex}{0.00873} &\raisebox{-1.0ex}{0.01636} 
&\raisebox{-1.0ex}{0.02243} &\raisebox{-1.0ex}{0.02758} &\raisebox{-1.0ex}{0.03503}
&\raisebox{-1.0ex}{Fig. \ref{op4I}}\\[1ex] 
\hline
 \raisebox{-1.0ex}{100}     &\raisebox{-1.0ex}{0.20471} &\raisebox{-1.0ex}{0.26402} 
&\raisebox{-1.0ex}{0.28301} &\raisebox{-1.0ex}{0.27541} &\raisebox{-1.0ex}{0.26570}
&\raisebox{-1.0ex}{Fig. \ref{op4II}}\\[1ex] 
\hline
 \raisebox{-1.0ex}{750}      &\raisebox{-1.0ex}{1.73811} &\raisebox{-1.0ex}{1.30137} 
&\raisebox{-1.0ex}{1.02320} &\raisebox{-1.0ex}{0.84397} &\raisebox{-1.0ex}{0.72334}
&\raisebox{-1.0ex}{Fig. \ref{op4III}}\\[1ex] 
\hline\hline
\end{tabular}
\caption{Average coalescence time for the total potential of Fig. \ref{pop4}}
\label{top4}
\end{table*}     


\begin{figure*}[th!]
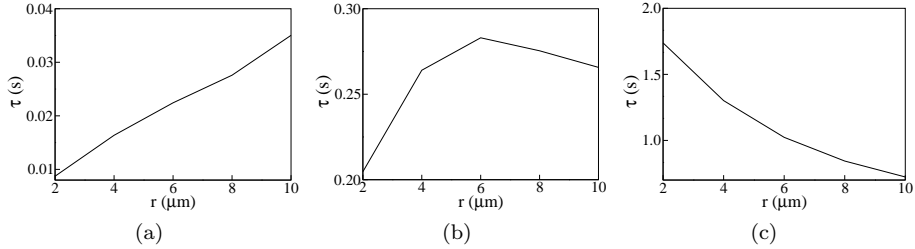

\bigskip
\begin{center}
\subfigure[]{
\label{op4I}
\includegraphics[scale=0.152]{Id20op4.eps}}
\subfigure[]{
\label{op4II}
\includegraphics[scale=0.152]{Id100op4.eps}}
\subfigure[]{
\label{op4III}
\includegraphics[scale=0.152]{Id750op4.eps}}
\bigskip
\caption{Dependence of the average coalescence time as a function of the particle radius for an initial distance of: (a) $d=20\,$nm, (b) $d=100\,$nm, (c) $d=750\,$nm. The total potential corresponds to the one depicted in Fig. \ref{pop4} (Table \ref{top4}).}
\end{center}
\end{figure*}

\begin{figure}[th!]
\centering
\includegraphics[angle=-90,scale=0.255]{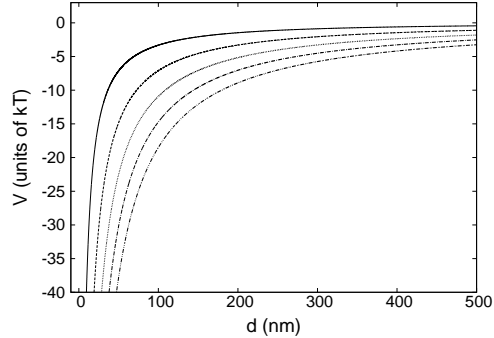}
\caption{Total potential of interaction between a large spherical droplet of $500\,\mu$m  and a small spherical droplet of micron-size. For this calculation the volume fraction of protein around the drops was assumed to be equal to $\bar{\phi}_\mathrm{p}=7.9\times 10^{-4}$. 
Solid line: $r=2\,\mu$m;
dashed line: $r=4\,\mu$m;
dotted line: $r=6\,\mu$m;
dot-dashed line: $r=8\,\mu$m; 
dashed double-dots line: $r=10\,\mu$m.}
\label{pop4}
\end{figure}

Table \ref{top4} and Figs. \ref{op4I}-\ref{op4III} show similar calculations in which the potentials depicted in Fig. \ref{pop4} are employed. These potentials were obtained, using a smaller volume fraction of protein ($\bar{\phi}_\mathrm{p}=7.9\times 10^{-4}$). This volume fraction produces a very small steric barrier ($\Delta V < 0.25$ kT) between $\kappa$-casein micelles. It also eliminates the repulsive barriers shown in Fig. \ref{pop3} for the drop-interface system, generating a total potential that is more attractive as the particle radius increases. Such potentials favor shorter coalescence times for larger particle radii. Hence, in this case, the attractive potential reinforces the effect of the buoyancy force and opposes the hydrodynamic resistance. 

Curiously, it is found that the coalescence time ($\tau$) increases with the size of the drops for an initial distance of $d=20\,$nm,. This calculation was repeated twice: first eliminating the Brownian movement of the particles, and second eliminating all repulsive potentials between them. However, similar results were found in all three cases. This indicated that the variation of $\tau$ with $r_i$ was due to the effect of the diffusion tensor. Table \ref{values} shows that the value of $f_\c^{(2)}$ for the Taylor tensor (Tensor $III$ in Table \ref{diffusion}) decreases sensibly as a function of the particle radius, promoting longer coalescence times for larger particle radius.

\begin{table}[th!] 
\centering
\begin{tabular}{cccc}
\hline\hline
$r_i$ ($\mu$m) &$4h_1r_i/r_*^2$ &$4h_2r_i/r_*^2$ &$4h_3r_i/r_*^2$\\ 
\hline\hline
2 & 0.0101 & 0.0504 & 0.3780 \\
4 & 0.0051 & 0.0254 & 0.1905\\
6 & 0.0034 & 0.0171 & 0.1280\\
8 & 0.0026 & 0.0129 & 0.0968\\
10& 0.0021 & 0.0104 & 0.0780\\
\hline\hline
\end{tabular}
\caption{The ratio $4h_ir_i/r_*^2$ for $h_1=20\,$nm, $h_2=100\,$nm and $h_3=750\,$nm.}
\label{values}
\end{table}    

As the initial distance of approach between the particles increases, the range of action of the attractive potential and the hydrodynamic friction increase. The diffusion tensor of Stokes depends linearly on the distance of approach, but inversely on the particle radius. According to Table \ref{values} the ratio $D_\t/D_0$ increases in absolute magnitude as the initial distance of approach increases. The particles diffuse more rapidly through a longer distance, and differences in the magnitude of the attractive potential are experienced by the drops during a longer period of time. As a result of this phenomenon, an increase in the initial distance of approach changes the variation of the coalescence time as a function of the particle radius. The interaction potential and the buoyancy force dominate at $d=750\,$nm, and the initial trend exhibited at $d=20\,$nm is reversed (see Table \ref{top4} and Fig. \ref{op4I}-\ref{op4II}). 

These results indicate that the dependence of the coalescence time as a function of the particle radius observed in the experiments requires: (a) the absence of repulsive barriers between the emerging drops and the interface, and (b) a sufficiently long distance of approach ($d\ge 750\,$nm). 

\begin{figure}[th!]
\centering
\includegraphics[angle=-90,scale=0.255]{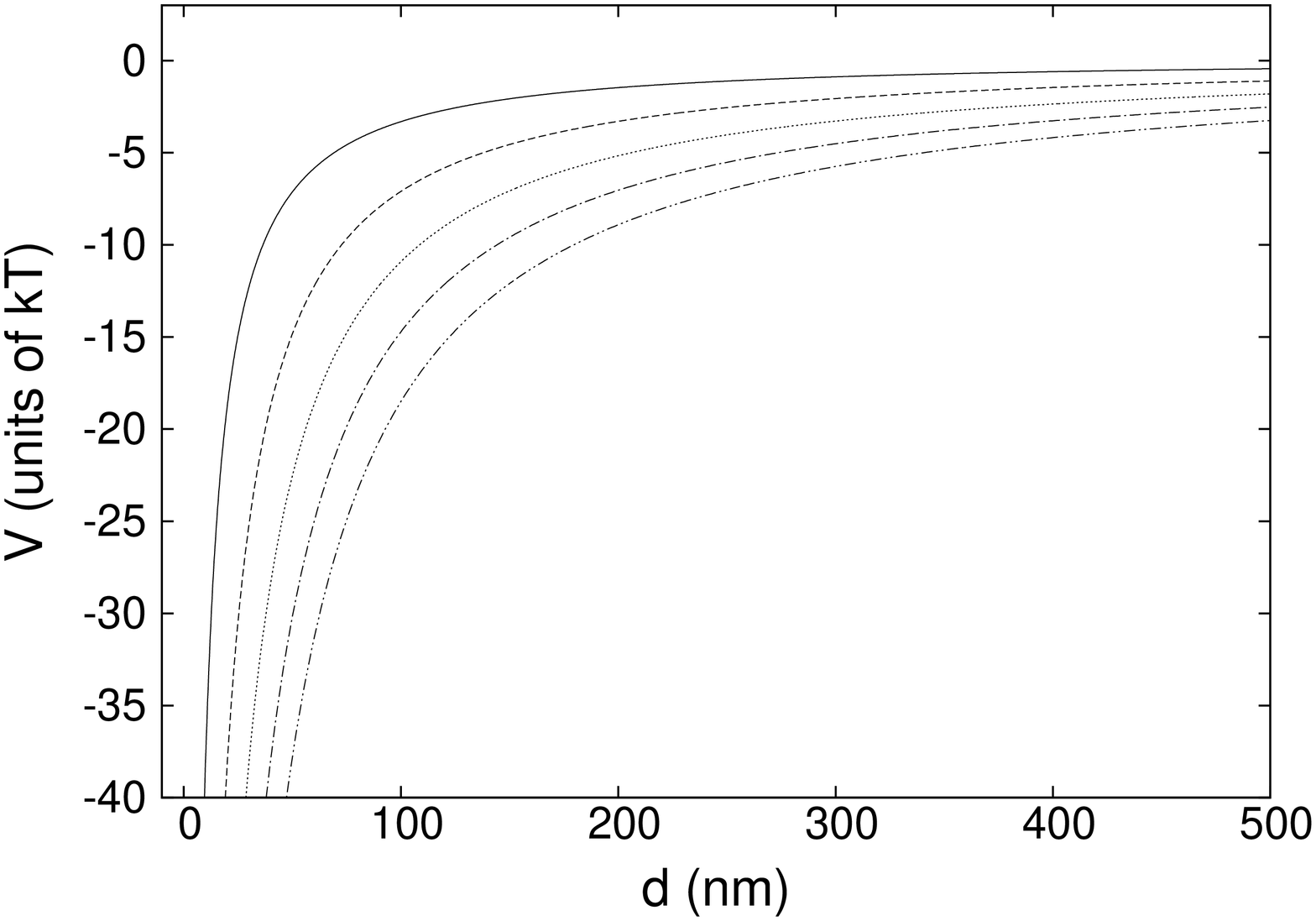}
\caption{Total potential of interaction between a large  deformable  droplet of $500\,\mu$m  and a small deformable droplet of micron-size. For this calculation the volume fraction of protein around the drops was assumed to be equal to $\bar{\phi}_\mathrm{p}=7.9\times 10^{-4}$. 
Solid line: $r=2\,\mu$m;
dashed line: $r=4\,\mu$m;
dotted line: $r=6\,\mu$m;
dot-dashed line: $r=8\,\mu$m; 
dashed double-dots line: $r=10\,\mu$m.}
\label{potdef}
\end{figure}

\begin{figure}[th!]
\centering
\includegraphics[scale=0.255]{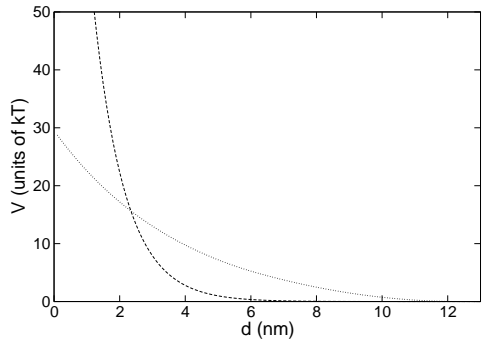}
\caption{Individual contributions to the total potential between a drop and a planar interface for a spherical droplet of $r=10\,\mu$m. The steric potential was calculated using a volume fraction of protein equal to $\bar{\phi}_\mathrm{p}=7.9\times 10^{-4}$. 
Dashed line: electrostactic;
dotted line: steric.}
\bigskip
\label{oneop4}
\end{figure}

\begin{figure}[th!]
\centering
\includegraphics[scale=0.255]{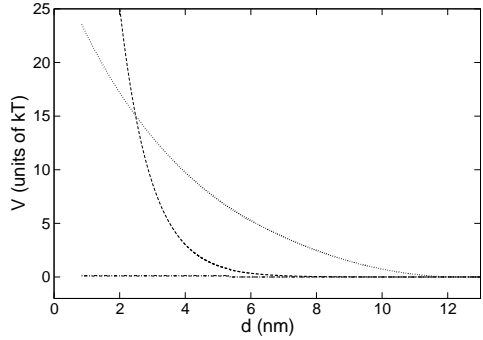}
\caption{Individual contributions to the total potential between a drop and a planar interface for a deformable droplet of $r=10\,\mu$m. The steric potential was calculated using a volume fraction of protein equal to $\bar{\phi}_\mathrm{p}=7.9\times 10^{-4}$. 
Dashed line: electrostactic;
dotted line: steric;
dot-dashed line: dilational; 
dashed double-dots line: bending.}
\label{onedef}
\end{figure}

Figures \ref{pop4} and \ref{potdef} show the final potentials employed in the present simulations for spherical and deformable drops. The total potential corresponding to each particle radius is completely attractive in both cases. The corresponding repulsive barriers illustrated in Figs. \ref{oneop4} and \ref{onedef} are counterbalanced by a large van der Waals attraction, see Fig. \ref{pop4} and \ref{potdef}. 

\begin{figure}[htpb]
\centering
\bigskip
\includegraphics[angle=-90,scale=0.275]{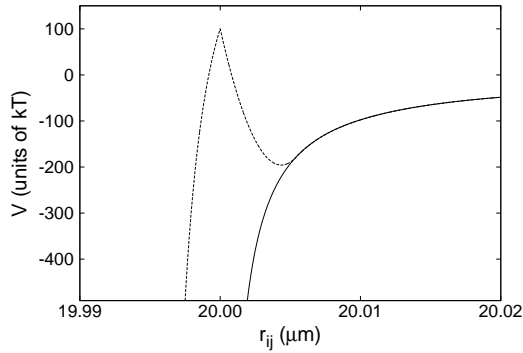}
\caption{Total interaction potential between two micron-size drops with a radius of $10\,\mu$m. The parameters of the potentials are equal to the ones employed in the calculation of Fig. \ref{pop4} and \ref{potdef} for the drop/interface model. $V_\mathrm{st}$ ($\bar{\phi}_\mathrm{p}=7.91\times 10^{-4}$). 
Solid line: spherical drops;
dashed line: deformable drops.}
\label{two}
\end{figure}

Figure \ref{two} shows the interaction potential between two drops of $r=10\,\mu$m. This potential was calculated using the same parameters that generate the potentials shown in Figs. \ref{pop4} and \ref{potdef} for the interaction between a drop and a planar interface. In the case of two deformable drops, the repulsive contributions produce a significant potential barrier, even at $\bar{\phi}_\mathrm{p}=7.9\times 10^{-4}$. Hence, care must be taken when extrapolating the experimental measurements of a drop/interface system to the coalescence behavior of two deformable drops.

Based on the results discussed above (Tables \ref{top3} and \ref{top4}), and the order of magnitude of the experimental coalescence times, several initial distances of approach were tested. For spherical particles d = $5\,\mu$m, $10\,\mu$m, $15\,\mu$m, $20\,\mu$m, and $30\,\mu$m. In the case of deformable drops, the order of magnitude of the experimental coalescence times is achieved at considerably shorter distances of separation. In this case initial distances $d=50\,$nm, $65\,$nm and $100\,$nm were found to be convenient. The results presented in the following figures correspond to those distances of approach that showed the closest agreement with the experimental measurements. 

All calculations included the effect of the buoyancy force, Eq. (\ref{bouyancy}). The time step of the simulations was kept large enough ($\Delta t_*= \Delta t \times D_0 /r_i^2=1.0$) in order to guarantee that $\Delta t>mD_0/kT$. For drops with $1\,\mu\textnormal{m}\le r_i \le 10\,\mu$m, $\Delta t_*=1$ corresponds to values between $2.0\times 10^{-6}\,$s and $2.0\times 10^{-2}\,$s, ($mD_0/kT=1.9\times 10^{-7}\,$s and $1.9\times 10^{-4}\,$s, respectively).

\section{Results and discussion}
\label{results}

\begin{figure}[th!]
\centering
\bigskip
\includegraphics[scale=0.255]{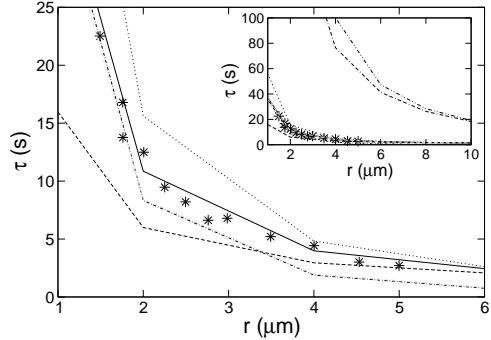}
\caption{Average coalescence time vs. droplet radius for spherical drops initially separated by a distance of $d=15\,\mu$m.
Stars: experimental data \cite{dickinson:1988};
double-dashed dot line: Tensor A;
dashed double-dots line: Tensor B;
dotted line: Tensor C;
grey dashed line: Tensor D;
solid line: Tensor E;
dashed line: Tensor F;
dot-dashed line: Tensor G; 
grey dot-dashed line: Tensor H
.}
\label{spherical}
\end{figure}

\begin{figure}[th!]
\centering
\bigskip
\includegraphics[scale=0.255]{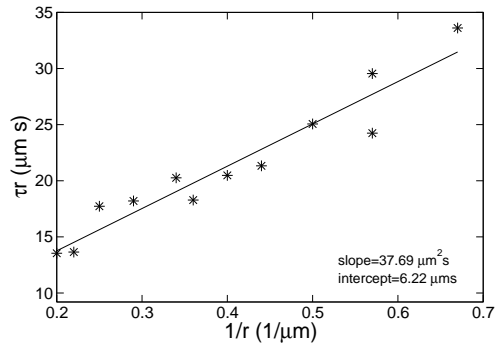}
\caption{Fitting of Eq. (\ref{Stokes-Taylor}) to the data of Dickinson {\it et al.} \cite{dickinson:1988}.
Stars: experimental data;
solid line: Stokes-Taylor equation}
\label{fit}
\end{figure}

Figure \ref{spherical} illustrates the results of the calculations for spherical drops. Notice that only six different particle radii were used in order to find the best combination of tensors. The separation distance which showed the closest agreement with the experimental data was $d=15\,\mu$m. This distance corresponds to the value of $h_\mathrm{ini}$ in Eq. (\ref{Stokes-Taylor}). Basheva {\it et al.} \cite{basheva:1999} suggested a simple procedure to estimate the distance $h_\mathrm{ini}$, at which spherical droplets change their bulk velocity drastically due to the proximity of an interface. According to Eq. (\ref{Stokes-Taylor}), a plot of the product $\tau r_i$ vs. $1/r_i$ should approach a straight line. Figure \ref{fit} shows the result of applying this procedure to the data of Dickinson {\it et al.} regarding hexadecane drops covered by $\beta$-casein. From the slope and the intercept of the curve, a value of $h_\mathrm{ini}$ equal to $d=19.4\,\mu$m was obtained. This value appears to be sufficiently close to $15\,\mu$m if it is considered that the procedure employed to implement the tensors in the simulations is not completely equivalent to the fitting of the analytical equation published in Ref. (\ref{st+ta}). 

According to Figure \ref{spherical}, all the simulations which include combinations of Stokes and Taylor tensors (Tensors D, E, G and H) come close to the experimental data. In the scale of this figure, Tensors G and H overlap, as well as tensors D and E. It appears from these results that the mobility of the liquid near the interface is not a very significant factor at least for this specific case. A close look at the curves evidences that Tensor D shows the best behavior. Tensor F underestimates the coalescence time, while Tensor C overestimates it. The behavior of Tensor F is partially due to the use of the Taylor tensor outside its region of validity. In regard to Tensor C it should be noticed that the quality of the prediction of the tensors which include the correction of Honig {\it et al.} markedly depend on the value of the reference radii employed. The equation of Honig {\it et al.} was formerly deduced for two equal spheres. This explains why a reasonable agreement is obtained with Tensor C, while the predictions of Tensors A and B are totally mistaken. In the last two cases, the radius of reference was estimated as the average radius between the emerging drop and the fixed $500\,\mu$m drop. As a result, $R_0$ lies far away from the radii of the coalescing drops, producing unreliable coalescence times.

\begin{figure}[th!]
\centering
\bigskip
\includegraphics[scale=0.255]{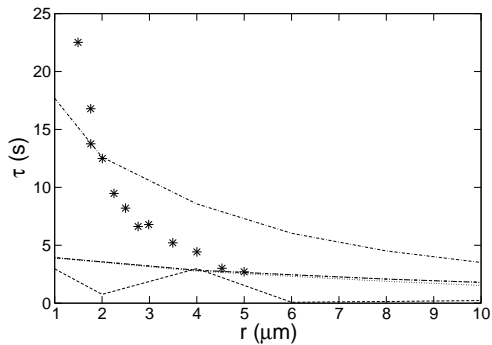}
\caption{Average coalescence time vs. droplet radius for deformable drops.
Stars: experimental data \cite{dickinson:1988};
dashed line: Tensor J ($d=50\,$nm);
dot-dashed line: Tensor K ($d=50\,$nm);
dotted line: Tensor L ($d=50\,$nm);
dashed double-dots line: Tensor $I$ ($d=50\,$nm);
double-dashed dot line: Tensor $I$ ($d=100\,$nm).} 
\label{deform}
\end{figure}

In Fig. \ref{deform} we show the behavior between average coalescence time and the droplet radius for the case of deformable drops. The value of $h_\mathrm{ini}$ was in all simulations of deformable droplets was of the order of 10 nanometers, three orders of magnitude different from the one deduced by application of Eq. (\ref{Stokes-Taylor}) to the experimental data (see Fig. \ref{fit}). Notice that one of the curves shows a non-monotonic variation of the coalescence time (Tensor J). This is due to the correction of the diffusion tensor of Reynolds when $f_\c^{(2)} > 1.0$. The program has the option to use the set of tensors available whatever their value, and also has the option to correct the calculated tensor using the expression of Honig {\it et al.}, if the absolute value of $f_\c^{(2)} > 1.0$. The fact that the correction was implemented in one of the simulations that used the  Reynolds tensor (Tensor J), indicates that analytical formula of this tensor might fall outside its range of validity when the scheme of Fig. \ref{ccfloc} is implemented.

The lowest monotonous curve in Fig. \ref{deform}, includes the calculations ran with an initial distance of $50\,$nm. These comprise Tensors I, L, and M. No significant difference was observed between the simulations ran with the tensor of Danov {\it et al.} at $\epsilon_\mathrm{S}=0.1$ and $1.0$. The highest monotonous curve include the calculations that employed an initial distance of $d=100\,$nm. The results from tensors I, K and M overlap in this case.  

In the case of Tensor K, all three mechanisms of coalescence previously described were tested. These included film drainage, and surface oscillations with two different forms of counting the time ($\tau_{ij}$). However, in the present case, the effect of the buoyancy force and the van der Waals attraction (deterministic forces) prevails. In the absence of a significant repulsive barrier, the drops coalesce once they enter the deformation zone. Hence, the evaluation of $\tau_{ij}$ is irrelevant regardless of the method of counting the coalescing time. Notice also, that the use of large time-steps in the simulations might induce the artificial coalescence of the particles since they might produce values of $\tau_{ij}$ considerably larger than $\tau_\mathrm{Vrij}$. However, the use of a large arbitrary time (i.e. $10\,$s) as input of the simulations led to the same result. Coalescence occurred very rapidly as a consequence of deterministic forces. This was further confirmed by setting the Brownian contribution of Eq. (\ref{brownian}) equal to zero. This is achieved using an auxiliary variable called �Damper� which multiplies the value of the random contributions. Notice that in the present simulations each particle moves individually with the equation of motion Eq. (\ref{brownian}). Hence, the random deviates are ascribed to each coalescing particle regardless the formation of a doublet. However, it is probable that during the drainage of the film, a couple of deformable drops might move as a unique entity. The program has the option to decrease the effect of the Brownian motion of each particle multiplying their random deviates by a real number lower than one.           

Tensors K and L were tested in order to consider the possible occurrence of deformation at higher distances of approach than the ones predicted by Eq. (\ref{hini}). The values of $h_\mathrm{ini}$ predicted by this equation do not surpass a few nanometers. However, as shown in Fig. \ref{two} for the case of non-deformable particles, the presence of a planar interface promotes a substantial viscous friction at distances of a few micrometers. Hence, the program has the possibility to introduce an arbitrary value of $h_\mathrm{ini}$ as an input of the calculation, but in this case the formula for the evaluation of the film radius cannot be used, and its value also needs to be estimated beforehand.

The methodology described for the simulation of deformable droplets is likely to be very useful in those cases in which the radius of the particles is larger, and a substantial repulsive potential exists. In the presence case, these procedures do not produce a better result than the simple combination of the tensors of Reynolds and Stokes (Tensor $I$). However, the fact that deformable droplets do not reproduce the experimental trend is not surprising. According to Ivanov {\it et al.} \cite{ivanov:1999} the analytic expression of $h_\mathrm{ini}$ in terms of the interaction forces and the disjoining pressure does not have a solution for droplet radii smaller than $83\,\mu$m. This means that the surface of the drop remains convex and there is no formation of a film. The present result support that prediction for radii between $1\,\mu\textnormal{m}\le r_i \le 10\,\mu$m.  

\begin{figure}[th!]
\centering
\bigskip

\includegraphics[scale=0.255]{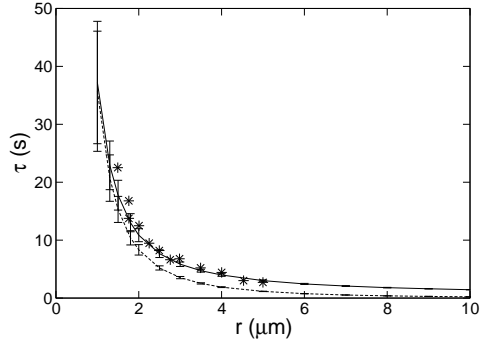}
\caption{Best simulation results for the coalescence time of a micron-size drop pressed by buoyancy against a planar interface (d=$15\,\mu$m).
Stars: experimental data \cite{dickinson:1988};
solid line: Stokes-Taylor law (Tensor D);
dashed line: Stokes-Taylor law (Tensor G).
Error bars were approximated by the standard deviation of $1000$ simulations}
\label{final}
\end{figure}

In order to make a closer comparison between the simulation and the experimental data, fifteen additional simulations corresponding to intermediate particle radii between $1$ and $10\,\mu$m were run for the case of spherical particles. For these calculations, the two tensors that produced the closest agreement with the experimental points were employed (Tensor D and G in Fig. \ref{spherical}).  Figure \ref{final} shows the results of the calculations. The agreement between the simulations and the experimental data is very good for the case of tensor D. Notice that the standard deviation of the calculations is indicated in the figure. It is remarkable that the error bars increase monotonically with the decrease of the particle radii. Such dependence was observed by Dickinson using lysozyme as a surfactant, but the magnitude of the errors was not plotted in Ref. \cite{dickinson:1988} for the case of the $\beta$-casein protein. It is likely that the error bars evidence the effect of the Brownian motion on the trajectory of the particles. However, a close analysis of the trajectories of each simulation was not carried out. 

The coalescence times predicted by Tensor G follow the curvature of the experimental points but do not get close enough. If the initial distance of separation is changed so that the experimental point corresponding to $r_i=4$ microns is reproduced, the rest of the coalescence times predicted lie above the experimental curve (for instance, $\tau=17.6\,$s for $r_i=2\,\mu$m, instead of $12.5\,$s as it was experimentally found). Moreover, the initial distance of separation has to be increased up to $30\,\mu$m in order to reproduce the point of $r_i=4\,\mu$m. This distance is considerably larger than the $19.2\,\mu$m deduced by application of Eq. (\ref{Stokes-Taylor}) to the experimental data of Dickinson et al. \cite{dickinson:1988}. Consequently, it appears that under the experimental conditions, the surface of the droplets acquire enough surfactant to behave as an immobile interface, despite the relatively short time of contact between the drops and the protein solution.

\section{Conclusion}
\label{conclusions}

In the absence of a significant repulsive barrier, the present simulations confirm the analytical  predictions of Basheva {\it et al.} \cite{basheva:1999}. According to our calculations, the experimental behavior of hexadecane drops at a water/hexadecane interface \cite{dickinson:1988} can be reproduced assuming spherical droplets that move with a combination of Stokes and Taylor tensors. The variation of the coalescence time as a function of the particle size, predicted by the simulations of deformable drops, does not follow the experimental trend. This suggests that micron size droplets behave as non deformable droplets in agreement with previous theoretical work \cite{ivanov:1997}.  

In was also shown that the presence of a strong repulsive barrier between the emerging drop and the interface might completely change the behavior of the coalescence time as a function of the particle radius. In this regard, the initial distance of approach between the particle and the interface plays a significant role.

\section*{Acknowledgements}

The authors acknowledge Dr. Aileen Losz\'an and Dr. Jhoan Toro-Mendoza for useful discussions and computational time. 

\bibliographystyle{apsrev}

\end{document}